\documentclass[12pt]{article}
\usepackage{amsmath}
\usepackage{graphicx}
\usepackage{float}
\usepackage{booktabs}
\usepackage{breqn}
\usepackage{makecell}
\usepackage{url} 

\newcommand{\blind}{0}

\usepackage{xcolor}

\usepackage[round]{natbib}

\begin{document}

\def\spacingset#1{\renewcommand{\baselinestretch}%
{#1}\small\normalsize} \spacingset{1}


\if0\blind

{
  \title{\bf A Covariate-Adjusted Homogeneity Test   with   Application to Facial Recognition Accuracy Assessment}

  \author{Ngoc-Ty Nguyen\\
    Department of Statistics and Data Science, University of Central Florida\\
    National Center of Forensic Science, University of Central Florida\\and\\
    P. Jonathon Phillips
     \\
     Information Access Division, National Institute of Standards and Technology\\and\\
     Larry Tang \\
    Department of Statistics and Data Science, University of Central Florida\\
    National Center of Forensic Science, University of Central Florida}
    \date{}
  \maketitle
} \fi

\if1\blind
{
  \bigskip
  \bigskip
  \bigskip
  \begin{center}
    {\LARGE\bf Title}
\end{center}
  \medskip
} \fi

\bigskip
\begin{abstract}
Ordinal scores occur commonly in   medical imaging studies and in  black-box  forensic studies   \citep{Phillips:2018}. To assess the accuracy of  raters in the studies, one needs to estimate the receiver operating characteristic (ROC)  curve     while accounting for covariates of raters. 
In this paper, we propose a covariate-adjusted homogeneity test to determine    differences in accuracy among multiple  rater groups.  We derived the theoretical results of the proposed test and conducted extensive simulation studies to evaluate the finite sample performance of the proposed test.     Our proposed test is applied to a face recognition study to identify statistically significant differences among five participant groups.   

\end{abstract}

\noindent%
{\it Keywords:} Covariate-adjusted ROC curve, AUC, power \& sample size, ordinal data, facial recognition 
 
\spacingset{2} 

\section{Introduction}
\label{sec:intro}

The receiver operating characteristic (ROC) curve  is a graphical plot that demonstrates classification ability of a marker when its separation threshold is changed.
In spite of initially being created for accuracy in radar detection, it is used widely in psychology \citep{swets:1973, Ratcliff:1992,youngstrom:2014,aggarwal2020predicting}, biometrics \citep{obuchowski:1997, pepe2000interpretation, ross:2003,zhu2021order}, medicine \citep{hanley1982meaning, centor1991signal, pencina2008evaluating, dhaya2020deep, tang2:12}, computer science \citep{hand2001simple, ferri2002learning, goncalves2020deep}. The main goal of building a ROC curve is to quantitatively differentiate two groups such as diseased and non-diseased subjects in medical diagnosis, pairs of face images from different persons and   pairs of images from the same person in facial recognition.  

Generally, a ROC curve depicts a relationship of true positive rate (TPR or sensitivity) and false positive rate (FPR or 1-specificity). In medical imaging studies, TPR is  the percentage of images of  diseased patients classified correctly, and FPR is the portion of images of non-diseased patients mislabeled. In fingerprint and face recognition, TPR is the percentage of two biometric samples from the same person correctly classified as being from the same person. FPR is the percentage of two biometric samples from different people incorrectly classified as being from the same person.
Because varying the decision threshold will change both the TPR and FPR, we computed the entire ROC.

In some settings, ordinal scores are preferred; for example, medical imaging \citep{wanyonyi2014image, gorham2010impact, toledano1999generalized} and facial recognition \citep{Phillips:2018}. \cite{wanyonyi2014image} used a 6-point criteria for crown–rump length measurements. \cite{gorham2010impact}  employed a 1–4 scale to score X-ray tubes. In \cite{toledano1999generalized}, a 5-point scale was utilized for staging lung cancer. Recently, \cite{Phillips:2018} applied 7-point scale to show raters' confidence whether two images are from one person in facial recognition. To address ordinal data, 
\citet{tosteson1988general} proposed a general regression method for estimating ROC curves from ordinal rating data. 

In numerous studies, researchers measure the effect of the levels of categorical covariates. Categorical covariates could be the imaging test; for example, magnetic resonance (MR) imaging and  computed tomography (CT) \citep{tosteson1994roc}, or different institutions
\citep{toledano1999generalized}. \cite{barlow2004accuracy} described levels of experience of radiologists as a categorical covariate that effects to accuracy of screening mammography interpretation.   Influence of race and gender demographics on estimates of the
accuracy of facial recognition algorithm were reported in \cite{o2012demographic}. Recently, \citet{Phillips:2018} compared face identification accuracy among five  participant groups including forensic facial examiners, facial reviewers, super-recognizers, fingerprint examiners, and students. The participants were asked to provide ordinal-scale decision scores for image pairs based on their belief on whether the pairs belong to the same  or different sources. In that paper, only the areas under ROC curves (AUCs) were computed to obtain accuracy of participant groups. Comparing accuracy of multiple categorical levels is desired to   identify  any differences in accuracy among the categories  (participant groups).   

To address this   research question, a homogeneity test is needed for inference on covariate adjusted ordinal ROCs. The null hypothesis of the test states, there is no difference in accuracy among categories; alternative hypothesis asserts, there is a difference. Unlike the test for AUCs \citep{tang2:12}, a test for ROC curves provides a complete picture of error rates for all possible thresholds.
In our method, ROC curves are built within ordinal regression framework in which rater groups or rater titles are considered as independent categories. This way, we obtain a  smooth ROC curve for each rater group and its uncertainties. By conducting the homogeneity test, the differences among categorical levels are tested. The regions of the ROC curves where the differences occur can be identified. Furthermore, within the homogeneity test context, the relationship between sample size and power is investigated. Determination of the minimum number of subjects for a test plays an important role because it helps to save resources such as labor, time, and cost. However, that relationship is not always explicit due to the complex nature of tests. In the present paper, 
the power and the minimum size are connected through the covariance matrix of estimated ROC curve.

The rest of the paper is arranged as follows. In Section \ref{sec:Homo_test}, a homogeneity test for ROC curves  is introduced.  The test statistic is based on estimated ROC curves from ordinal regression.  Section \ref{sec:theory} presents the theoretical results of the proposed test.   Section 4 introduces our results of power and sample size analysis. In Section \ref{sec:Simu}  we carry out  extensive simulations studies with various numbers of rater groups and evaluate the final sample performance regarding Type I error rates and powers.  The proposed test   is applied for the aforementioned face recognition study to compare examiner groups  in Section \ref{sec:face}. We characterized differences among the five participant groups. The differences identified by our method are consistent with those in \citet{Phillips:2018}. In addition,  our method gives the regions of ROC curves in which  the differences in examiner groups occur.  This is an important contribution of the proposed method since higher matching accuracy at low FPRs is of interest in  operational decisions. The conclusion is in Section \ref{sec:conc}.
\section{Covariate-Adjusted Homogeneity Test}
\label{sec:Homo_test}
\subsection{Notations}

In the present paper, we denote the upper-case letter X as data which contains covariates as columns and samples as rows. $X=x$ is then understood as a specific value of covariates in the data. In the data, upper-case $R$ is used to describe the ordinal score whose values are from $1$ to $L$ where $L$ is called ordinal scale. Also, a binary status is denoted as $D$ where $D=1$ or $D=0$ splits observations into two sub-classes. Upper-case $G$ is the number of rater groups who give ordinal scores as assessing subjects.

In this paper, we use ROC curve to characterize the accuracy of performances. Let Y denote a continuous random variable related to scores in evaluation.
The general formula of a ROC curve is expressed as a function of FPR as 
	 \begin{equation}\label{gen_ROC}
	 ROC\left(t \right) =S_{1}\left( S_{0}^{-1}\left( t\right) \right),\quad t\in\left( 0,1\right)  
	 \end{equation}
where $S_0\left(c\right)=P\left(Y \ge c \vert D=0\right)$ and $S_1\left(c\right)=P\left(Y \ge c \vert D=1\right)$ are FPR and TPR with threshold $c$.  If $S_{0}$ and $S_{1}$ follow normal distributions, equation \eqref{gen_ROC} is expressed as
	 \begin{equation}\label{bi_ROC}
	 ROC\left(t \right) =\Phi\left( \dfrac{\mu_{1}-\mu_{0}}{\sigma_{1}}+\dfrac{\sigma_{0}}{\sigma_{1}}\Phi^{-1} \left( t\right) \right),\quad  t\in\left( 0,1\right)  
	 \end{equation}
where $\mu_{0}, \mu_{1}, \sigma_{0},\sigma_{1}$ are the means and the standard deviations of two sub-populations, respectively. Then, the AUC also has a explicit form as $AUC=\Phi\left((\mu_1-\mu_0)/\sqrt{\sigma_{1}^2+\sigma_{0}^2}\right)$.
In the present study, $ROC_{\textbf{x},g}(t)$ and  $AUC_{\textbf{x},g}$  are the ROC curve and the corresponding AUC at a specific covariates $\textbf{x}$ of group $g$ with $g=1, \cdots, G$. Covariates in face recognition data are raters' group, age, gender. Also, because ROC curves are built within framework of the ordinal regression, their variances are determined by variance of parameters in the model. For the sake of making inference easily, we denote $\widehat{ROC}_{\textbf{x},g}\left(t\right)$ and $\widehat{AUC}_{\textbf{x},g}$ be the estimated ROC curve and AUC at covariate $\textbf{x}$ of group $g$ and estimated parameter $\hat{{\boldsymbol\gamma}}$ of the ordinal regression.             
\subsection{A Homogeneity Test}

In this section, we introduce a homogeneity test for ROC curves. 
Assume that there are $G$ rater groups each of which includes $J_1, J_2, \cdots, J_G$ members assessing $K=K_0+K_1$ subjects such as images in medical diagnostics or image pairs in fingerprint or facial recognition. Out of $K$, there are $K_0$ non-diseased subjects in  medical diagnostic or $K_0$ different sources image pairs in fingerprint or facial recognition and $K_1$ diseased or same source ones. Accuracies of groups are characterized by $ROC_{x,1}(t), \cdots,ROC_{x,G}(t)$ or $AUC_{x,1}, \cdots, AUC_{x,G}$ respectively. The goal is to test homogeneity among groups. The null hypothesis of the test is stated as all groups have the same accuracy while the alternative is supported if there exist differences among groups. It is noteworthy to mention that the ROC curves are functions of TPR with respect to FPR. Therefore, the test  is conducted at each fixed FPR.  
We define a vector as
\begin{equation*}
\boldsymbol\Lambda=\left( ROC_{x,1}(t), ROC_{x,2}(t),\cdots,ROC_{x,G}(t)  \right)^{\prime}. 
\end{equation*}
If a new vector $\boldsymbol\Lambda_{C}$ is defined by subtracting $ROC_{x,G}(t)$ from $\boldsymbol\Lambda$ as
\begin{equation*}
\boldsymbol\Lambda_{C}=\left( ROC_{x,1}(t)-ROC_{x,G}(t), \cdots,ROC_{x,G-1}(t)-ROC_{x,G}(t)  \right)^{\prime},
\end{equation*}
the null hypothesis is now formulated as $H_{0}:\boldsymbol\Lambda_{C}=\textbf{0}$ vs. $H_{a}:\boldsymbol\Lambda_{C}\neq \textbf{0}.$
The relationship between $\boldsymbol\Lambda_{C}$  and $\boldsymbol\Lambda$ can be expressed as $\boldsymbol\Lambda_{C}=\mathcal{K}\boldsymbol\Lambda$ where $\mathcal{K}=\left(I_{G-1},-\textbf1_{G-1}\right)$ with an identity matrix $I_{G-1}$ and a vector of one's $\textbf1_{G-1}$. In this case, we use $ROC_{G}(t)$ as a reference for comparison purpose. In fact, any group can be in charge of the role.

With a given data, we need to estimate the ROC curves to proceed the test. An estimate of a ROC curve, denoted with a hat, can be retrieved nonparametrically or parametrically \citep{zhang2012linear}. With the first approach, an empirical ROC curve is obtained. Alternatively, parametric methods need to assume distributional forms for two populations. Thus, ROC curve can be derived analytically. In this paper, we use the later technique with binormality assumption for scores within framework of ordinal regression discussed in the next section. 
\subsection{ROC Estimators based on Ordinal ROC Regression}
\label{sec:est} 
Assume that we want to bridge $L$-scale ordinal scores $R$ with observable variables comprised in a matrix $X$. Without loss of generality, we denote the first column of $X$ as $D$ which is a binary variable of 0 or 1. Then, $D$ splits observations into two sub-groups such as diseased and non-diseased status in medical diagnostics, genuine and imposter scores in facial recognition. 

We use a location-scale model to  estimate the covariate-specific   ROC curve. 
 In the model, each of outcomes $R_i, i=1,2,\cdots N$ links to an example which is described by a vector of covariates, $X_i=\left\lbrace D_i,x_{i1},...,x_{ip} \right\rbrace $ or $\left\lbrace D_i,\textbf{x}_i\right\rbrace$ where  $p$ is the number of covariates and N is the total number of observations. It is noteworthy that out of $p$ covariates, one represents for group status of raters.
The ordinal ROC regression starts by supposing that discrete outcomes $R$ belong to a latent continuous variable $T$ which can be partitioned into sub-regions by thresholds $ -\infty = \tau_{0} < \tau_{1} <\cdots <\tau_{L-1}< \tau_{L}=\infty$.  The outcome $R$ receives the value $r_{l}$ if $\tau_{l-1}<T \leq \tau_{l}$.
	 The general formula of ordinal regression can be expressed as 
	 \begin{equation*}\label{eq_ordinal}
	g\left[  \phi_{l}\left( R \le l\vert\textbf{x}\right) \right] =\frac{\tau_{l}-\left( \alpha_{0}D+\boldsymbol\alpha_{1}\textbf{x}+D\boldsymbol\alpha_{2}\textbf{x}\right)}{\exp\left(\beta_{0}D+\boldsymbol\beta_{1}\textbf{x}+D\boldsymbol\beta_{2}\textbf{x} \right) }, 
	 \end{equation*}
	 where $l=1,2,...,L-1$ and $\textbf{x}$ denote for any of $\lbrace{\textbf{x}_i \rbrace}_{i=1}^N$ where $N=K\sum_{g=1}^{G}J_{g}$ is the total number of observations, $g\left( \cdot \right) $ is the link function, $\phi_{l}\left( R \le l\vert\textbf{x}\right)$ is the cumulative probability that $R \le l $, a vector production, for example $\boldsymbol{\alpha}_1\textbf{x}$, is written as $\boldsymbol{\alpha}_1\textbf{x} = \alpha_{11}x_{1}+\cdots+\alpha_{1p}x_{p}$.  
	 With the probit link, the model is rewritten as  
  \begin{equation*}\label{probit}
	\phi_{l}\left( R \le l\vert\textbf{x}\right) =\Phi\left\lbrace \frac{\tau_{l}-\left( \alpha_{0}D+\boldsymbol\alpha_{1}\textbf{x}+D\boldsymbol\alpha_{2}\textbf{x}\right)}{\exp\left(\beta_{0}D+\boldsymbol\beta_{1}\textbf{x}+D\boldsymbol\beta_{2}\textbf{x} \right) }\right\rbrace ,   
	 \end{equation*}
	 where $\Phi\left(\cdot \right) $ is the standard normal cumulative distribution function. With this approach, the latent variables for a particular covariate $\textbf{x}$ are normally distributed with means and standard deviations described in Table \ref{tb:Ord_ROC}.
\begin{table} [H]
\centering
\caption{Ordinal Regression ROC Parameters}
\begin{tabular}{|l|c|c|}
\hline
\thead{} & \thead{$D=1$ }& \thead {$D=0$}
        \\ \cline{1-3}
\hline
Mean & $\mu_1=\alpha_0+\left(\boldsymbol{\alpha}_1+\boldsymbol{\alpha}_2\right)\textbf{x}$ & $\mu_0=\boldsymbol{\alpha}_1\textbf{x}$\\\hline
Standard dev. &$\sigma_1=\exp\lbrace{ \beta_0+\left(\boldsymbol{\beta}_1+\boldsymbol{\beta}_2\right)\textbf{x} \rbrace}$ & $\sigma_0=\exp\lbrace{\boldsymbol{\beta}_1\textbf{x}\rbrace}$\\
\hline 
 \end{tabular}  \label{tb:Ord_ROC}
\end{table}
Substitute the means and standard deviations in Table \ref{tb:Ord_ROC} into (\ref{bi_ROC}), the ROC curve within the framework of ordinal regression for a specific covariate $\textbf{x}$ of group $g$  is finalized as   
	 \begin{equation}\label{ord_ROC}
	 ROC_{\textbf{x},g}\left(t  \right) =\Phi\left( \frac{\alpha_{0}+\boldsymbol\alpha_{2}\textbf{x}}{\exp\left(\beta_{0}+\boldsymbol\beta_{1}\textbf{x}+\boldsymbol\beta_{2}\textbf{x}\right)}+\frac{1}{\exp\left(\beta_{0}+\boldsymbol\beta_{2}\textbf{x}\right)}\Phi^{-1} \left( t\right) \right), \quad  t\in\left( 0,1\right) . 
	 \end{equation}
The corresponding $AUC$ is also expressed as 
	 \begin{equation}\label{ord_AUC}
	 AUC_{\textbf{x},g} =\Phi\left( \frac{\alpha_{0}+\boldsymbol\alpha_{2}\textbf{x}}{\sqrt{\exp\left(2\beta_{0}+2\boldsymbol\beta_{1}\textbf{x}+2\boldsymbol\beta_{2}\textbf{x}\right)+\exp\left(2\boldsymbol\beta_{1}\textbf{x}\right)}}\right ). 
	 \end{equation}
Obviously, ROC curve in \eqref{ord_ROC} and AUC in \eqref{ord_AUC} are determined by coefficients which are estimated by maximizing the likelihood function. To simplify notations, we use $\boldsymbol\gamma$ as a composite form of thresholds $\tau_{l}$, coefficients $\alpha_{0},\boldsymbol\alpha_{1},\boldsymbol\alpha_{2},\beta_{0},\boldsymbol\beta_{1},\boldsymbol\beta_{2}$ and $\hat{\boldsymbol\gamma}$ is its estimator, i.e. $\hat{\boldsymbol\gamma} \equiv \left(\hat{\boldsymbol{\tau}}, \hat{\alpha}_{0},\hat{\boldsymbol\alpha}_{1},\hat{\boldsymbol\alpha}_{2},\hat{\beta}_{0},\hat{\boldsymbol\beta}_{1},\hat{\boldsymbol\beta}_{2}\right)$. 
\section{Theoretical Results of   the Proposed Homogeneity Test}\label{sec:theory}
In this  section, we investigate asymptotic distribution of ROC curves and corresponding AUCs. We then study the asymptotic results of the proposed test. Based on the results, the sample size and power analysis is conducted. 
\subsection{Theoretical Property of ROC estimators based on Ordinal Regression
}We start with asymptotic normality of the maximum likelihood estimator of parameters $\boldsymbol\gamma \equiv \left(\boldsymbol{\tau}, \alpha_{0},\boldsymbol\alpha_{1},\boldsymbol\alpha_{2},\beta_{0},\boldsymbol\beta_{1},\boldsymbol\beta_{2}\right)$.\\
 \textbf{Lemma 1}: 
Let $R_1, R_2,\cdots R_N$ be independent  with density $f\left(R_i\vert \boldsymbol\gamma\right)$ and the pairs $\lbrace R_i,X_i \rbrace_{i=1}^N$ to be $i.i.d.$. If 
 \begin{itemize}
     \item $\Theta$ is an open and convex in $\Re^k$ where $k$ is the dimension of $\boldsymbol\gamma$,
     \item 
     $\sum_{i=1}^N X_iX_i^{\prime}$ has full rank,
     \item $\frac{n_{g,D=1}}{n_{g,D=0}} \rightarrow \lambda_g >0 $ as $n_{g,D=1} \rightarrow \infty$ and $n_{g,D=0} \rightarrow \infty$, i.e. the ratio of two sub-classes converges to a constant that is greater than 0 for all rater groups,
     \item
    $l_N\left(\boldsymbol\gamma\right)=\sum_{i=1}^{N}l_i \left(\boldsymbol\gamma\right)$ is the log likelihood function and concave,
     \item $H_N\left(\boldsymbol\gamma\right)=\sum_{i=1}^{N}H_{i}=-\sum_{i=1}^{N}\partial^2l_i\left(\boldsymbol\gamma\right)/\partial\boldsymbol\gamma\partial\boldsymbol\gamma^{\prime}$ and
     $\max_{\boldsymbol\gamma \in \aleph_N\left(\delta\right)}\Vert V_{N}\left(\boldsymbol\gamma\right)-\boldsymbol I\Vert \rightarrow 0$ 
     \\
     where  $V_{N}\left(\boldsymbol\gamma\right)=H_N^{-1/2}\left(\boldsymbol{\gamma}_0\right)H_N\left(\boldsymbol{\gamma}\right)H_N^{-T/2}\left(\boldsymbol{\gamma}_0\right)$
     \\ and
     $\aleph_N\left(\delta\right)
     =\lbrace \boldsymbol\gamma: \Vert H_N^{T/2}\left(\boldsymbol{\gamma}_0\right)\left(\boldsymbol\gamma -\boldsymbol\gamma_0\right)\Vert \leq\delta \rbrace$ where $\boldsymbol{\gamma}_0$ is the true value of  $\boldsymbol{\gamma}$,
     \item solution of $\partial l_N\left(\boldsymbol\gamma\right)/\partial\boldsymbol\gamma=0$ exists and denoted as $\hat{\boldsymbol\gamma}$,
 \end{itemize}
 then
 
 $\sqrt{N}\left(\hat{\boldsymbol\gamma} - \boldsymbol\gamma_0\right) \xrightarrow[]{d}N\left(\boldsymbol0,\Sigma_{\boldsymbol{\gamma}_0}\right)$
 
 where
 $\Sigma_{\boldsymbol{\gamma}_0}=E\left[H\left(\boldsymbol\gamma_0\right)\right]^{-1}$
 
The proof of Lemma 1 is sketched in  Appendix.  

We make following assumptions to derive asymptotic properties of estimated ROC curves and AUCs. 
\begin{itemize}
 \item Assumption 1: We assume that $ROC_{\textbf{x}, g}\left(t\right)$ is differential, and hence continuous, with respect to $\boldsymbol\gamma$ for all $g=1, \cdots, G$. 
\item Assumption 2: We assume that $AUC_{\textbf{x},g}$ is differential, and hence continuous, with respect to $\boldsymbol\gamma$ for all $g=1, \cdots, G$. 
\end{itemize}
\textbf{Theorem 1}:  Under Assumption 1, 
   $ \sqrt{N}\left[\widehat{ ROC}_{\textbf{x},g}\left(t\right) - ROC_{\textbf{x},g}\left(t\right)\right] $ converges in distribution to a normal zero-mean random variable with variance  $V_{ROC_{\textbf{x},g}\left( t\right)}=J_{ROC_{\textbf{x},g}\left( t\right)}\Sigma_{\boldsymbol{\gamma}_0} J^{\prime}_{ROC_{\textbf{x},g}\left( t\right)}$
where $J_{ROC_{\textbf{x},g}\left( t\right)}=\nabla_{\boldsymbol{\gamma}} ROC_{\textbf{x},g}\left( t\right)\vert_{\boldsymbol{\gamma}=\boldsymbol{\gamma}_0}$ as $N \rightarrow \infty$ for all $g=1, \cdots, G$.  

Proof:
With Assumption 1, expanding $\widehat{ROC}_{\textbf{x},g}\left(t\right)$ about $\boldsymbol{\gamma}_0$  yields to 
\begin{equation}
\label{eq:exp_ROC}
      \widehat{ROC}_{\textbf{x},g}\left(t\right)=ROC_{\textbf{x},g}\left(t\right)+J_{ROC_{\textbf{x},g}\left(t\right)}\left(\hat{\boldsymbol\gamma}-\boldsymbol\gamma_0\right)+o\left(\left(\hat{\boldsymbol\gamma}-\boldsymbol\gamma_0\right)^2\right).
\end{equation}
Due to $ \sqrt{N}\left( \hat{\boldsymbol\gamma} - \boldsymbol\gamma_0\right) \xrightarrow[]{d}N\left(\boldsymbol0,\Sigma_{\boldsymbol{\gamma}_0}\right)$ in Lemma 1 and $o\left(\left(\hat{\boldsymbol\gamma}-\boldsymbol\gamma_0\right)^2\right) \xrightarrow[]{p} \textbf{0}$ as $N  \xrightarrow[]{} \infty$, we obtain
\begin{equation*}
\sqrt{N}\left[ \widehat{ROC}_{\textbf{x},g}\left(t\right) - ROC_{\textbf{x},g}\left(t\right)\right] \xrightarrow[]{d} N\left(0,J_{ROC_{\textbf{x},g}\left( t\right)}\Sigma_{\boldsymbol{\gamma}_0} J^{\prime}_{ROC_{\textbf{x},g}\left( t\right)}\right)     .
\end{equation*}
\textbf{Corollary 1}: Under Assumption $2$, 
   $\sqrt{N} \left[ \widehat{AUC}_{\textbf{x},g} -  AUC_{\textbf{x},g}\right] $ converges in distribution to a normal zero-mean random variable with variance $V_{AUC_{\textbf{x},g}}=J_{AUC_{\textbf{x},g}}\Sigma_{\boldsymbol{\gamma}_0} J^{\prime}_{AUC_{\textbf{x},g}}$
where $J_{AUC_{\textbf{x},g}}=\nabla_{\boldsymbol{\gamma}} AUC_{\textbf{x},g}\vert_{\boldsymbol{\gamma}=\boldsymbol{\gamma}_0}$ as $N \rightarrow \infty$ for all $g=1, \cdots, G$.  

Proof: 
Integrating both sides of Equation \ref{eq:exp_ROC} over $t$ in Theorem 1, we have
\begin{equation*}
      \widehat{AUC}_{\textbf{x},g}=AUC_{\textbf{x},g}+J_{AUC_{\textbf{x},g}}\left(\hat{\boldsymbol\gamma}-\boldsymbol\gamma_0\right)+o\left(\left(\hat{\boldsymbol\gamma}-\boldsymbol\gamma_0\right)^2\right).
\end{equation*}
Here we use Assumption 2 that $J_{AUC_{\textbf{x},g}}$ exists.
Due to $ \sqrt{N}\left( \hat{\boldsymbol\gamma} - \boldsymbol\gamma_0\right) \xrightarrow[]{d}N\left(\boldsymbol0,\Sigma_{\boldsymbol{\gamma}_0}\right)$ in Lemma 1 and $o\left(\left(\hat{\boldsymbol\gamma}-\boldsymbol\gamma_0\right)^2\right) \xrightarrow[]{p} \textbf{0}$ as $N  \xrightarrow[]{} \infty$, we obtain
\begin{equation*}
\sqrt{N}\left[ \widehat{AUC}_{\textbf{x},g} - AUC_{\textbf{x},g}\right] \xrightarrow[]{d} N\left(0,J_{AUC_{\textbf{x},g}}\Sigma_{\boldsymbol{\gamma}_0} J^{\prime}_{AUC_{\textbf{x},g}}\right)     .
\end{equation*} 
\subsection{Asymptotic Property of the Proposed Test}
Let $\hat{\boldsymbol\Lambda},\hat{\boldsymbol\Lambda}_C$ be estimators of $\boldsymbol\Lambda$ and $\boldsymbol\Lambda_{C}$ which can be written as
$
   \hat{\boldsymbol\Lambda}=\left( \widehat{ROC}_{\textbf{x},1}\left(t\right), \cdots,\widehat{ROC}_{\textbf{x},G}\left(t\right)\right)^{\prime},$
   and $\hat{\boldsymbol\Lambda}_{C}=\left(\widehat{ROC}_{\textbf{x},1}\left(t\right)-\widehat{ROC}_{\textbf{x},G}\left(t\right),\cdots ,\widehat{ROC}_{\textbf{x},G-1}\left(t\right)-\widehat{ROC}_{\textbf{x},G}\left(t\right)\left(t\right) \right)^{\prime}.$
The relationship $\hat{\boldsymbol\Lambda}_C=\mathcal{K}\hat{\boldsymbol\Lambda}$ still holds for estimators. 
The test statistic is defined as 
\begin{equation*}
\label{test_stat}
\Psi=\widehat{\boldsymbol\Lambda}_{C}\left(\text{Var}\widehat{\boldsymbol\Lambda}_{C}\right)^{-1}\widehat{\boldsymbol\Lambda}_{C}^{\prime}.
\end{equation*}
The variance $\text{Var}\widehat{\boldsymbol\Lambda}_{C}$ is dependent on covariance matrix $\Sigma_{\hat{\boldsymbol{\gamma}}}$ which is asymptotically approximated as $\Sigma_{\boldsymbol{\gamma}_0}$. Using \eqref{eq:exp_ROC}  and
concatenating all $ \widehat{ROC}_{\textbf{x},g}\left(t\right)$ for $g=1,\cdots, G$ yields to   
\begin{equation*}
    \hat{\Lambda}=\Lambda+\text{F}\left(\hat{\boldsymbol\gamma}-\boldsymbol\gamma_0\right)+o\left(\left(\hat{\boldsymbol\gamma}-\boldsymbol\gamma_0\right)^2\right),
\end{equation*}
where $F=\left(J_{ROC_{\textbf{x},1}\left(t\right)},\cdots,J_{ROC_{\textbf{x},G}\left(t\right)} \right)^{\prime}$. 
Using  $\hat{\boldsymbol\Lambda}_C=K\hat{\boldsymbol\Lambda}$ and taking variance both sides yields to
\begin{equation*}
\label{Var_Lamda}
    \text{Var}\hat{\boldsymbol\Lambda}_{C}=\text{KF}\Sigma_{\hat{\boldsymbol{\gamma}}}\text{F}^{\prime}\text{K}^{\prime},
\end{equation*}
where $\Sigma_{\hat{\boldsymbol{\gamma}}}$ depends on the sample size. Employing Lemma 1, one can see that $\hat{{\Lambda}}_C$ asymptotically follows a multinormal distribution given by $N_{G-1}\left(\Lambda_{C},\Sigma_{G-1}=KF\Sigma_{\boldsymbol\gamma_0} F^{\prime}K^{\prime}\right)$.  \\
\textbf{Theorem 2}: If conditions in Lemma 1 and Assumption 1 are satisfied, under the null hypothesis, $\Psi$ converges in distribution to a Chi-square distribution with $G-1$ degrees of freedom $\chi^2_{G-1}$ and under the alternative, $\Psi$ still converges to a Chi-square distribution with the same degrees of freedom but with a non-centrality parameter $\eta=\boldsymbol{\Lambda}_{C}\left(\text{Var}\boldsymbol{\Lambda}_{C}\right)^{-1}{ {\boldsymbol\Lambda}}_{C}^{\prime}$ as $N \xrightarrow{} \infty$.

 The proof of Theorem 2 can be found in  Appendix.  

Given a significance level $\alpha$, the null hypothesis is rejected if $\Psi>\chi^2_{G-1,\alpha}$ where $\chi^2_{G-1,\alpha}$ is the critical value of a Chi-square distribution with $G-1$ degrees of freedom.

Determining non-centrality parameter occurs in various statistical analysis, such as the analysis of variance for tests of homogeneity, Chi squared test for goodness of fit, power analysis. Since the power analysis usually relates to the sample size problem, the non-centrality  parameter can be used to determine the minimum sample size provided the power is supplied .

Solution of a power problem replies on the availability of the non-centrality of a Chi-squared distribution. Early, \citet{Haynam1970} prepared tables for the non-centrality parameter of a Chi-squared distribution with some given values of degree of freedom, significance level and power. Then, \citet{guenther1977power} calculated the minimum sample size  for the three most frequently used tests at given power using those tables. Next, \citet{saxena1982estimation} estimated the non-centrality parameter of a Chi-squared distribution by employing the maximum likelihood technique. However, only were the lower and upper bounds derived instead of a closed form for the parameter. Thanks to developing of computer technology, nowadays we can numerically compute the non-centrality parameter. 

The power $1-\beta$ where $\beta$ is the probability of a type II error is defined as 
\begin{equation}
\label{eq_typeII}
    1-\beta=P\left(\chi^2_{G-1}\left(\eta\right) > \chi^2_{G-1,\alpha}\right).
\end{equation}
With given values of $\alpha,\beta$ and $G$, the non-centrality parameter $\eta$  can be determined by solving \eqref{eq_typeII}.  Denote $\eta_{\beta,\alpha}$ be the solution of \eqref{eq_typeII}, using definition of $\eta$ yields to
\begin{equation}
\label{eq_samplesize}
    \eta_{\beta,\alpha}= \boldsymbol\Lambda_{C}\left(\text{KF}\Sigma_{\gamma_0}\text{F}^{\prime}\text{K}^{\prime}\right)^{-1}\boldsymbol\Lambda_{C}^{\prime}.
\end{equation}
The minimum sample size which is fewest samples that satisfies equality \eqref{eq_samplesize} is determined by using Equation \eqref{eq_typeII} to obtain $\eta_{\alpha,\beta}$ numerically and scanning sample size until the equality  \eqref{eq_samplesize} is satisfied. Although Equation \eqref{eq_samplesize} is derived for ROC curves, a similar one is also obtained for AUCs. 

\section{Simulation Studies}
\label{sec:Simu}
\subsection{Simulation Settings}
In this part, we describe our design for simulation. Our data includes  ordinal scores, a continuous variable $X_1$ and discrete covariates representing for rater groups. The latent scores of $g^{th}$ group  are normally distributed as
\begin{equation*}
\label{ord_sim}
 \begin{gathered}[b]
   T_{g}\vert \left(X_{1},D=1\right) \sim N\left(1+2X_{1}+\psi+a_{g},\phi\text{Var}\left(e_{1}\right) \right),\\
   T_{g}\vert \left(X_{1},D=0\right) \sim N\left(1+X_{1},\text{Var}\left(e_{0}\right) \right),
   \end{gathered}
\end{equation*}
where $X_1$ is uniformly distributed in $\left[0,1\right]$, $e_{0}, e_{1}$ are standard normal distributions. In those equations , parameters $\psi$, $a_g$ and $\phi$ control the distance and difference in variances between two normal distributions. Moreover, parameter $a_{g}$, which only depends on the group label is utilized to adjust differences in ROC curves among groups. On the other hand, parameter $a_g$ is able to control the null and alternative hypothesis. With those distributions, the true ROC curve and the corresponding AUC are expressed as
\begin{equation}
    \begin{gathered}
      ROC_{\textbf{x},g}^{true}\left(t\right) =\Phi\left( \frac{x_1+\psi+a_g}{\sqrt{\phi}}+\frac{1}{\sqrt{\phi}}\Phi^{-1} \left( t\right) \right) \quad  t\in\left( 0,1\right),
      \\
      AUC_{\textbf{x},g}^{true}=\Phi\left(\frac{x_1+\psi+a_g}{\sqrt{1+\phi}}\right).
    \end{gathered}
\end{equation}
By adjusting the value of $a_g$, we end up with four settings in Table \ref{tb:Setting}. 
\begin{table} [H]
\caption{Settings for simulation study }\centering
\begin{tabular}{|l|l|l|}
\hline
\thead{Setting} & \thead{Description}& \thead {$a_g$}
        \\
        \hline
\thead{1} & \thead{G identical groups }& \thead { $a_{g}=0, \quad g=1,\cdots,G$}
        \\
        
\thead{2} & \thead{G-1 identical groups, \\ the last one with a higher accuracy }& \thead { $a_{g}=0, \quad g=1,\cdots,G-1$ \\ and $a_G=0.5$}
        \\
\thead{3} & \thead{G-2 identical groups, \\ two last ones with higher accuracies}& \thead {$a_{g}=0, \quad g=1,\cdots,G-2$ \\ and $a_{G-1}=a_G=0.5$}
        \\

\thead{4} & \thead{G groups have different accuracies}& \thead {$a_{1}=0, a_{g+1}-a_{g}=0.2 \quad g=1,\cdots,G-1$}
        \\
        
\hline
 \end{tabular}  \label{tb:Setting}
\end{table}

\begin{figure}[H]
  \begin{center}
\includegraphics[height = 0.31\textheight]{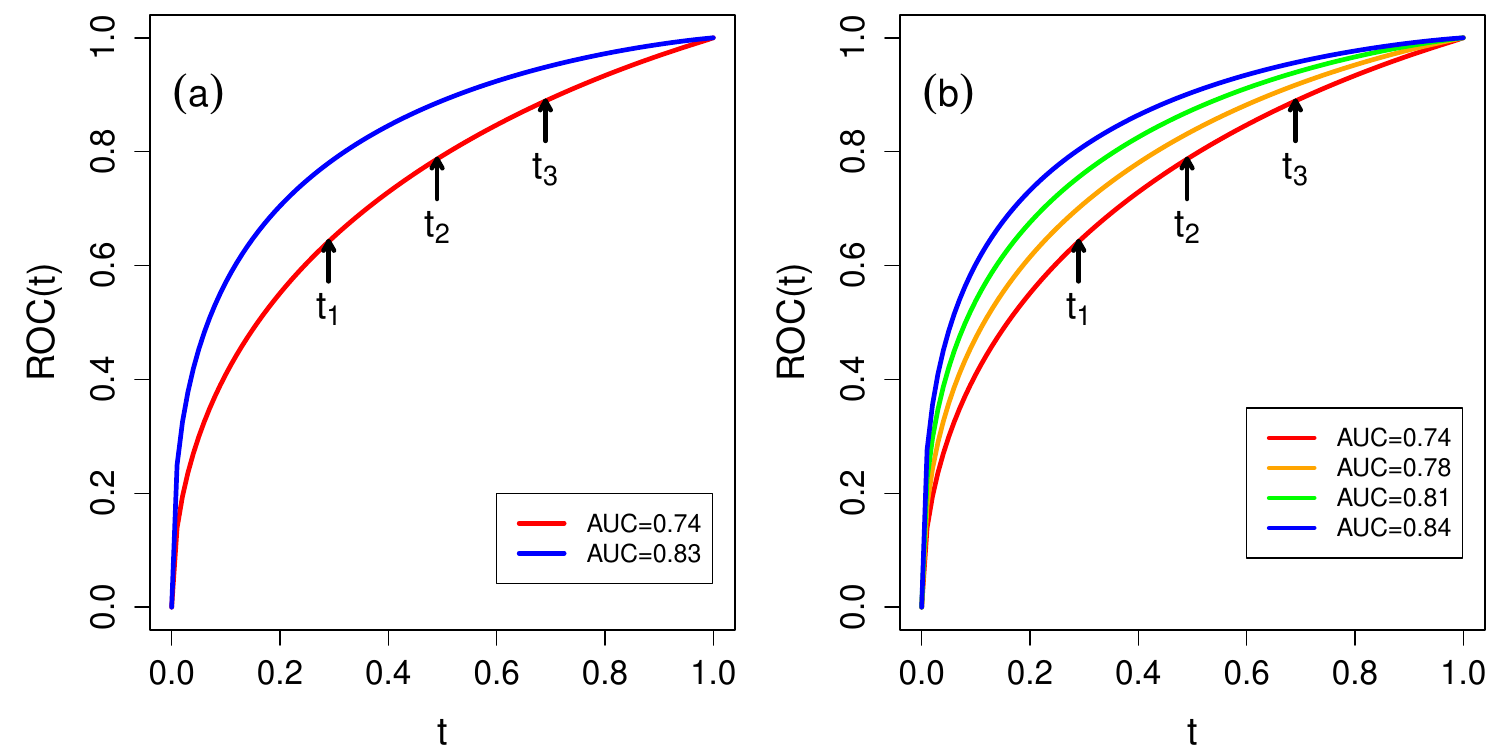}
\end{center}
\caption{ROC curves for null and alternative hypothesises. Arrows point positions used for the test. Numbers in legend are AUCs corresponding with colored curves. G=4, L=7 and  $\psi=0.5$, and $\phi=1.5$, $x_{1}=0.5$ are used.}
\label{fig:hypo}
\end{figure}

Among four settings, the first one supports for the null hypothesis of the homogeneity test while others describe alternative ones. In the Figure \ref{fig:hypo}, we illustrate settings with four groups. In Fig.\ref{fig:hypo}a, the red curve are the identical ROC curve if groups have the same accuracy. If there exists one or two groups with higher accuracy, the blue ROC curve appears (setting 1 or 2). If groups have different accuracies, four ROC curves are described as in Fig.\ref{fig:hypo}b. In Fig.\ref{fig:hypo}, arrows are also added at FPRs of $0.3,0.5,0.7$ on the curves where values are used in the homogeneity tests. 
    
\subsection{Numerical Results}

\subsubsection{Consistency}
\begin{figure}[H]
  \begin{center}
\includegraphics[height = 0.32\textheight]{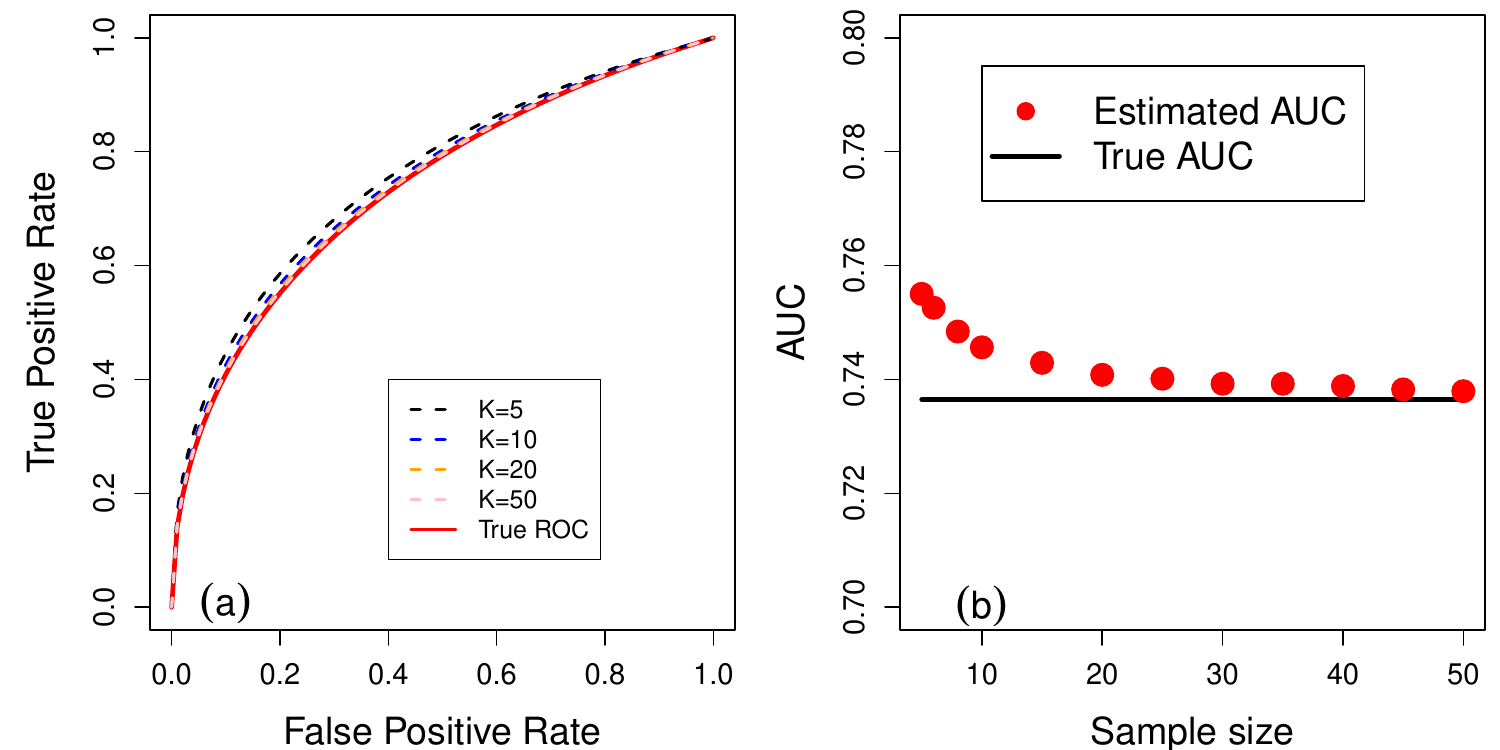}
\end{center}
\caption{Convergence of  ROC curves, AUCs when the sample size changes. G=5, J=10, L=7 and  $\psi=0.5$, and $\phi=1.5$, $x_{1}=0.5$ are used. (a): Estimated ROC curves for four selected sample sizes of 5 (dashed black), and 10 (dashed blue), 20 (dashed orange) and 50 (dahsed pink). Red solid  line is the true ROC curve. (b): Dots are estimated AUCs and black solid line is the exact AUC of 0.736.}
\label{fig:cvg}
\end{figure}

First, we examine the consistency of estimated ROC curves and AUCs.  We assume that all groups have the same number of members, i.e. $J_1=J_2=\cdots=J_G=J$. Hence, term "sample size" K should be understood as the number of samples assigned to each rater. Setting 1 with 10000 data sets are simulated in this subsection. In Figure \ref{fig:cvg}, estimated ROC curves and AUCs are depicted with some sample sizes where $G=5$ and $L=7$ are used. Value of other parameters can be seen in the caption.  
In Fig.\ref{fig:cvg}, estimated ROC curves of four selected sample sizes are depicted along with the true curve. It is obvious that the larger the sample size is, the closer the estimated curve approaches the exact one. In this case, a sample size between 10 and 20 is a reasonably optimal value leading to a consistent ROC curve. 

Next, in Fig.\ref{fig:cvg}b, estimated AUCs with sample sizes varying from 5 to 50 is presented. One can see that estimated AUC asymptotically converges to the true value. Indeed, the bias of estimators are just around 2\% at sample size of 5 and less than 1\% at 15. Thus, 15 could be consider as asymptotic value of the sample size.      

Futhermore, we validate the quality of estimators and their variance by calculating the confidence interval coverages of difference in ROC curves and AUCs. Let $\Delta ROC_{12}\left(t\right)=ROC_2\left(t\right)-ROC_1\left(t\right), \Delta AUC_{12}=AUC_2-AUC_1 $ be the difference in ROC curves or in AUCs between the second and the first group. The coverage of $\Delta ROC_{12}\left(t\right)$ curves is the portion of the 10000 curves bounded by the $\left(100-\alpha\right)\%$ confidence interval that is

\begin{figure}[H]
  \begin{center}
\includegraphics[height = 0.32\textheight]{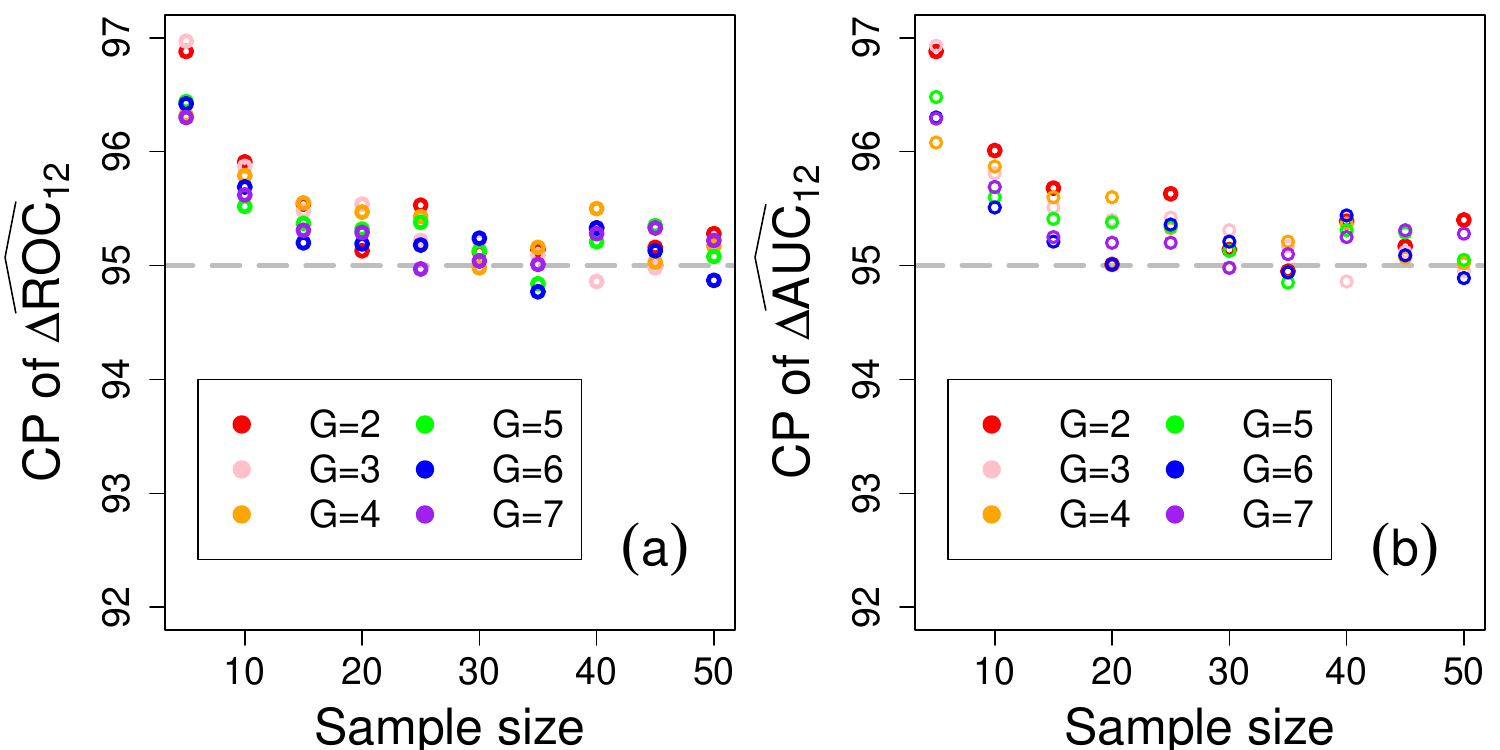}
\end{center}
\caption{Coverage probabilities of the 95\% confidence intervals of $\Delta \widehat{ROC}_{12}$ curve at FPR of $0.3$ (a) and of $\Delta\widehat{ AUC}_{12}$ (b). The dashed grey line is the nominal level. J=10, L=7 and  $\psi=0.5$, and $\phi=1.5$, $x_{1}=0.5$ are used.}
\label{fig:cicvg}
\end{figure}

$\left(\Delta\widehat{ ROC}_{12}\left(t\right)-Z_{\frac{\alpha}{2}}\times\sqrt{\text{var}\left(\Delta\widehat{ ROC}_{12}\left(t\right)\right)},\Delta\widehat{ ROC}_{12}\left(t\right)+Z_{\frac{\alpha}{2}}\times\sqrt{\text{var}\left(\Delta\widehat{ ROC}_{12}\left(t\right)\right)}\right).$ 
Similarly, coverage of $\Delta AUC_{12}$ is bounded by\\
$\left(\Delta\widehat{ AUC}_{12}-Z_{\frac{\alpha}{2}}\times\sqrt{\text{var}\left(\Delta\widehat{ AUC}_{12}\right)},\Delta\widehat{ AUC}_{12}+Z_{\frac{\alpha}{2}}\times\sqrt{\text{var}\left(\Delta\widehat{ AUC}_{12}\right)}\right).$ 
The coverages of $\Delta ROC_{12} \left( t \right)$ at $t_1=0.3$ and $\Delta AUC_{12}$ with different number of groups and sample sizes are presented in Figure \ref{fig:cicvg}.

In Fig.\ref{fig:cicvg}a, confidence interval coverage of $\Delta ROC_{12}$ at FPR of $0.3$ is illustrated and those of $\Delta AUC_{12}$ are shown in Fig.\ref{fig:cicvg}b. In both figures, one can see that the portions approach to 95\% starting from the sample size of 100 and get closer when the size increases regardless of the number of groups. It is noticeable that the result for $\Delta ROC_{12}$ is  presented at one value of FPR but the similar ones are also obtained at different points on the ROC curves. It implies that the convergence occurs for the entire $\Delta ROC_{12}$. Moreover, results for other pairs are also found analogous to that of $\Delta ROC_{12}$.     

\begin{figure}[H]
  \begin{center}
\includegraphics[height = 0.32\textheight]{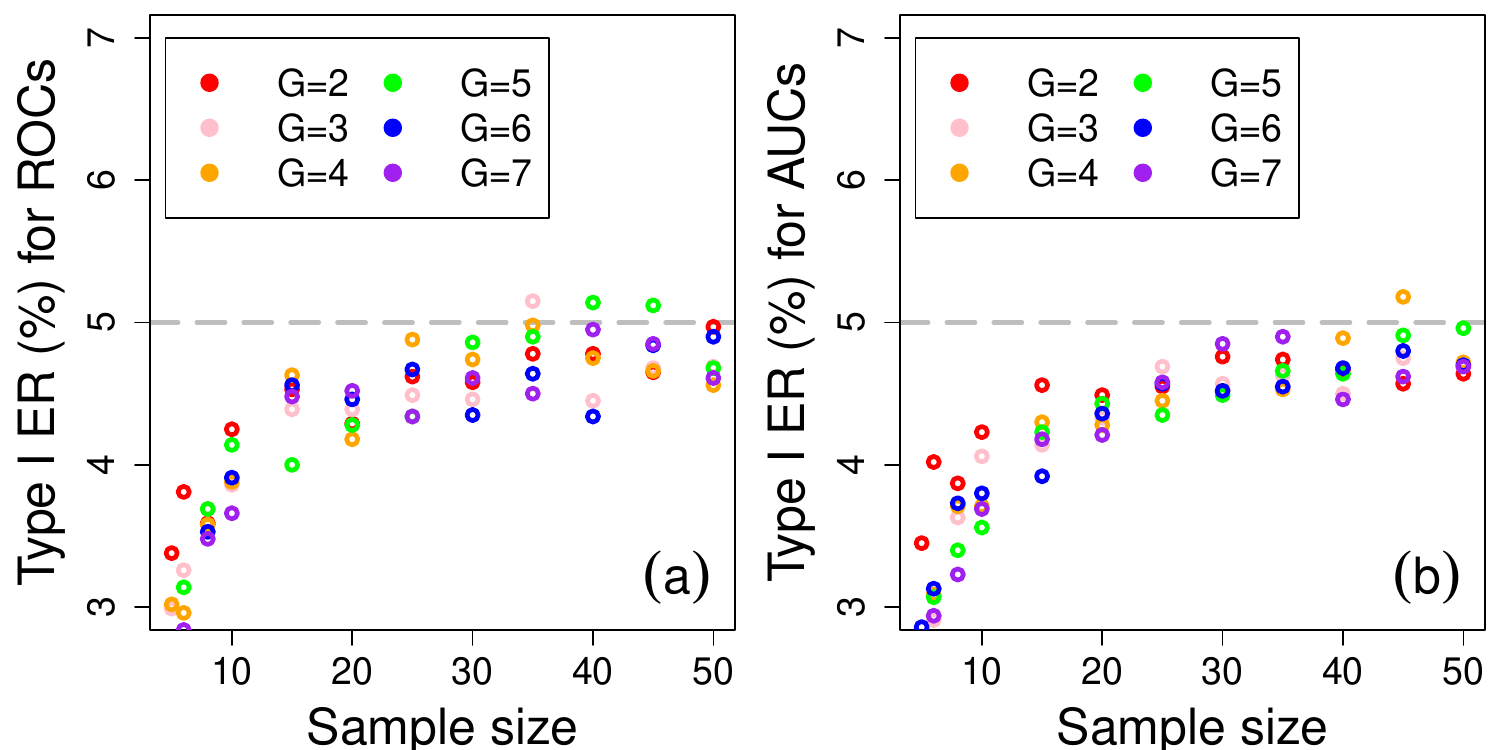}
\end{center}
\caption{Type I error rate for the test for ROCs at FPR of $0.3$ (a) and for AUCs (b). The dashed grey line is 5\% significance level. Parameters are fixed the same as in Figure \ref{fig:cicvg}.}
\label{fig:typeI}
\end{figure}

In addition, we compute the probability of Type I error of the homogeneity test. In
Figure \ref{fig:typeI}, we depict Type I error rate of the test for ROC curves at FPR of $0.3$ (a) and for AUCs on (b). 
As seen in Fig.\ref{fig:typeI}, regardless of the number of group, the Type I error approaches 5\% for both of tests using ROC curves and AUCs. Analogous results are also retrieved for different values of FPR.  

 \subsubsection{Minimum Sample Size}
 \label{sec:sample_size}
 In this subsection, we illustrate calculation of minimum sample sizes given $\alpha$ and $\beta$ with settings supporting for the alternative hypothesis. By using Equation \eqref{eq_typeII} to obtain $\eta_{\alpha,\beta}$ numerically and scanning sample size until equality in \eqref{eq_samplesize} is satisfied, we determine the minimum sample size for each setting.  In this subsection, assumption $J_1=J_2=\cdots=J_G=J$ is used first but unequal number of raters are also investigated latter. Values of FPR on ROC curves $\lbrace 0.3,0.5,0.7\rbrace$ used for calculation  is marked in Fig.\ref{fig:hypo}. \\
 First, with evenly distributed samples,  minimum sample sizes to reach a probability of a Type I error of 5\% and a power of 80\% is demonstrated in Table \ref{tab:Mini_size}. With each setting, the minimum sample size is calculated by using TPR at three different FPRs, denoted at $t_1,t_2,t_3$. Estimated sample size using AUCs is also provided. With setting 2, a vector of AUC values is $\Lambda=(0.736,0.736,\cdots, 0.736, 0.829)$ whose last entry is larger than others. With setting 3, $\Lambda=(0.736,0.736 ..., 0.736, 0.829, 0.829)$ where last two identical elements are larger than others. With setting 4 where all elements are different, $\Lambda=(0.736, 0.776, 0.812, 0.844, 0.873, 0.897, 0.918)$ for seven groups. If less groups are needed, for instance five groups, first five elements are used.              
 \begin{table}[H] 
\centering
\caption{Minimum sample sizes with $\alpha=0.05,\beta=0.2$. J=10, L=7 and  $\psi=0.5$, and $\phi=1.5$, $x_{1}=0.5$ are used. FPRs at $t_1=0.3, t_2=0.5, t_3=0.7$ are selected. }
\begin{tabular}{ccccccccccccc}

\hline
\hline
      &     \multicolumn{4}{c}{Setting 2} & \multicolumn{4}{c}{Setting 3} & \multicolumn{4}{c}{Setting 4}\\
\cmidrule(lr){2-5}
\cmidrule(lr){2-5}
\cmidrule(lr){6-9}
\cmidrule(lr){6-9}
\cmidrule(lr){6-9}
\cmidrule(lr){10-13}
\cmidrule(lr){10-13}

G&$t_1$  &$t_2$&$t_3$&$AUC$ &$t_1$  &$t_2$&$t_3$&$AUC$ &$t_1$  &$t_2$&$t_3$&$AUC$
\\ 

\rule{0pt}{15pt}
3	&41	&47	&58	&41	&42	&49	&68	&43	&80	&89	&115	&80	\\
4	&35	&37	&47	&35	&30	&35	&46	&30	&37	&43	&56	&37	\\
5	&33	&35	&41	&33	&24	&28	&34	&24	&19	&24	&30	&22	\\
6	&31	&34	&38	&31	&21	&23	&30	&25	&12	&14	&19	&12	\\
7	&29	&31	&36	&29	&20	&22	&26	&20	&8	&10	&13	&8	\\
\hline
\hline
\end{tabular}
 \label{tab:Mini_size}
\end{table}

As seen in Table \ref{tab:Mini_size}, with each setting, a larger sample size is needed if a higher FPR is used. This can be explained as at higher FPR, the gap between curves are narrower. That leads to an increase in sample size to reduce the variance because the non-centrality parameter on left side of Equation \eqref{eq_samplesize} is fixed. Furthermore, with three settings, the minimum sample sizes retrieved from ROC curves at FPR of 0.3 are similar to those from AUCs. That could be because at FPR of $0.3$, the gaps in TPRs, $\boldsymbol\Lambda_{C}^{\lbrace{ROC\rbrace}}$ and difference in AUCs, $\boldsymbol\Lambda_{C}^{\lbrace{AUC\rbrace}}$, among rater groups are similar. 

Next, we investigate the scenario in which groups have different number of raters. Denote $n_1:n_2:\cdots:n_G$ be the ratio of the number of raters among groups, i.e. $J_1=n_1J, \cdots, J_G=n_GJ$. 
We use setting 4 with four groups for following calculations. The minimum sample size presented in Table \ref{tab:Mini_size2} are the number of samples assigned for each rater. As seen in the table, the minimum sample size is sensitive to changes of the ratio. Assume that among four groups, one has twice samples than others which is described in first four rows in Table \ref{tab:Mini_size2}. It is obvious that which group has more samples influences the total minimum sample size. This finding is also seen with different ratios of samples. Once again, the minimum sample sizes retrieved from higher FPRs are larger than that from a lower one. Additionally, sample sizes obtained by using FPRs of 0.3 is still similar to those from AUCs.       
 \begin{table}[H]
\centering
\caption{Minimum sample sizes with different number of raters in groups. $\alpha=0.05,\beta=0.2$ J=10, L=7 and  $\psi=0.5$, and $\phi=1.5$, $x_{1}=0.5$ are used. FPRs at $t_1=0.3, t_2=0.5, t_3=0.7$ are selected.}
\begin{tabular}{cccccccccc}
\hline
\hline
Ratio &$t_1$  &$t_2$&$t_3$ & AUC & Ratio &$t_1$  &$t_2$&$t_3$ & AUC \\
\hline
\hline
1:1:1:2	&28	&33	&45	&29	&2:1:1:2	&20	&23	&30	&20	\\
1:1:2:1	&37	&42	&57	&37	&2:1:2:1	&25	&28	&38	&25	\\
1:2:1:1	&35	&39	&51	&35	&1:2:3:4	&20	&24	&33	&20	\\
2:1:1:1	&26	&29	&38	&26	&4:3:2:1	&18	&19	&24	&18	\\
1:1:2:2	&29	&32	&46	&29	&4:2:2:1	&17	&19	&25	&17	\\
1:2:1:2	&26	&30	&40	&26	&1:2:2:4	&20	&24	&32	&20	\\
2:2:1:1	&25	&28	&37	&25	&4:2:1:4	&10	&11	&15	&10	\\
1:2:2:1	&34	&37	&50	&34	&4:1:2:4	&10	&12	&16	&10	\\

\hline
\hline

\end{tabular}
 \label{tab:Mini_size2}

\end{table}

\section{Application to Facial Recognition Accuracy}
\label{sec:face}

Forensic facial examiners perform detailed comparisons between images of two faces and determine if the faces are from the same person or different people. Examiners' extensive training and qualifications allow them to give expert opinion in court proceedings.  Because of facial examiners' detailed comparisons, the field of  facial forensics   is  a pattern-based forensic discipline. Two reports identified the necessity to empirically measure error rates for pattern-based disciplines in forensics \citep{national2009,pcast2016report}. \citet{Phillips:2018} provided the needed scientific evidence of facial examiners' ability by conducting a study that measured examiners' accuracy when they performed forensic comparisons. To assess examiners' ability relative to other groups, the study measured the accuracy of forensic facial reviewers, super-recognizers, fingerprint examiners, and students. Forensic facial reviewers are trained to perform facial comparisons faster than examiners. Super-recognizers possess a natural ability to recognize faces. Fingerprint examiners specializing in comparing latent fingerprints. Students served as a proxy for the general population. 

Next we give an overview of the methods in \citet{Phillips:2018}. The participants consisted of 57 facial examiners, 30 facial reviewers, 13 super-recognizers, 53 fingerprint examiners and 31 students. Each participate judged the similarity of the same 20 face-pairs.  For each face-pair, participants judged the similarity of the two faces on a 7-point scale, with +3 for the highest confidence of same person to –3 for the highest confidence of different people. 

\citet{Phillips:2018} computed accuracy at the individual level by computing the AUC for each participant. They reported overall group accuracy with the median AUC of the group and compared two groups with the Mann-Whitney test. In our analysis, we pool participants for each of the five groups and we assume members within the same group have the same accuracy. Using scores as outcomes and group status as covariates, we estimate the ROC curves and the corresponding AUCs for each group. The two methods produce slightly different results, but overall the results from the two studies are consistent.  

\subsection{Estimated ROC Curves and AUCs}

\begin{figure}
  \begin{center}
\includegraphics[height = 0.6\textheight]{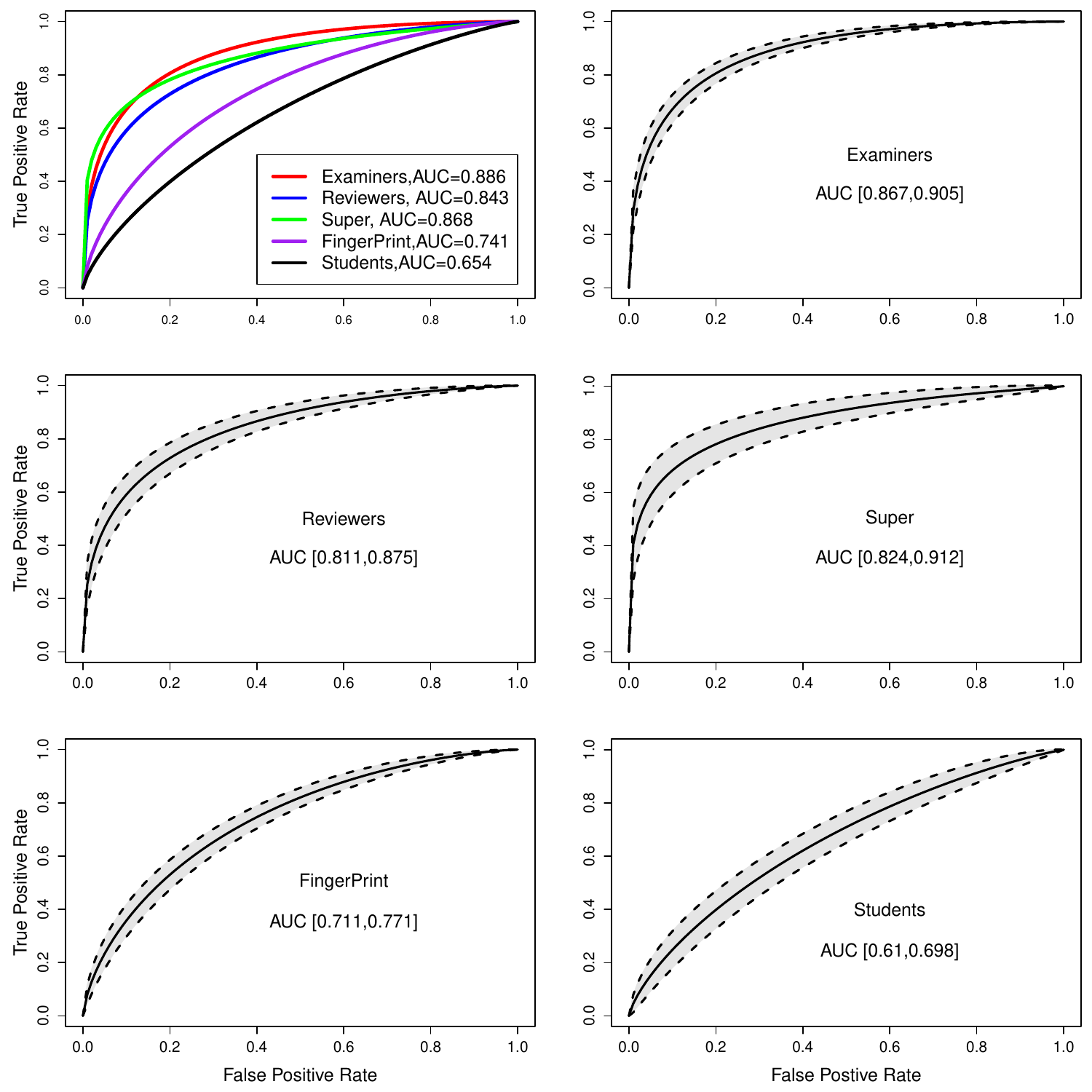}
\end{center}
\caption{Plot of estimated ROCs and their 95\% confidence bands for the five participant groups. The upper left panel shows the estimated ROCs for all five groups and the legend reports the estimated AUC for each group. The remaining panels plot the ROC and confidence bands for each group individually. In each panel, the solid lines is the estimated ROC, the dashed lines are upper and lower bound of the 95\% confidence bands, and the legend states the group and reports the 95\% lower and upper confidence intervals for the AUC.}
\label{fig:CI_ROC}
\end{figure}

We start our analysis by applying our ordinal regression technique to facial recognition ratings and estimating the ROCs and AUCs for each of the first subject groups. For the ROCs we compute the 95\% confidence bands and for the AUC we compute the 95\% confidence intervals.  Figure~\ref{fig:CI_ROC} shows estimated ROCs and AUCs with corresponding 95\% confidence bands and intervals for each group.  Based on the AUC estimates, the facial examiners has the highest AUC followed by  super-recognizers, facial reviewers, fingerprint examiners, and students. This order agrees with \citet{Phillips:2018}.

\subsection{Homogeneity Test}

Next, we check if the the AUCs and ROCs for five groups are statistically different. We formulate this question as a hypothesis test with the null hypothesis that the AUCs (respectively ROCs) for all five groups are statistically the same. If the null hypothesis is not true, then at least one of group's AUCs (respectively ROCs) are statistically different than other four groups. First, we test AUCs and  for the five groups, then the ROCs. 

For AUCs, we compute the homogeneity test statistic $\Psi=129.2$. Since this test statistic is larger than $\chi^2\left(0.95,df=4\right)=9.49$, the null hypothesis is rejected with a 95\% confidence level, and the AUCs are not the same for all five groups. 

\begin{figure}
  \begin{center}
\includegraphics[height = 0.35\textheight]{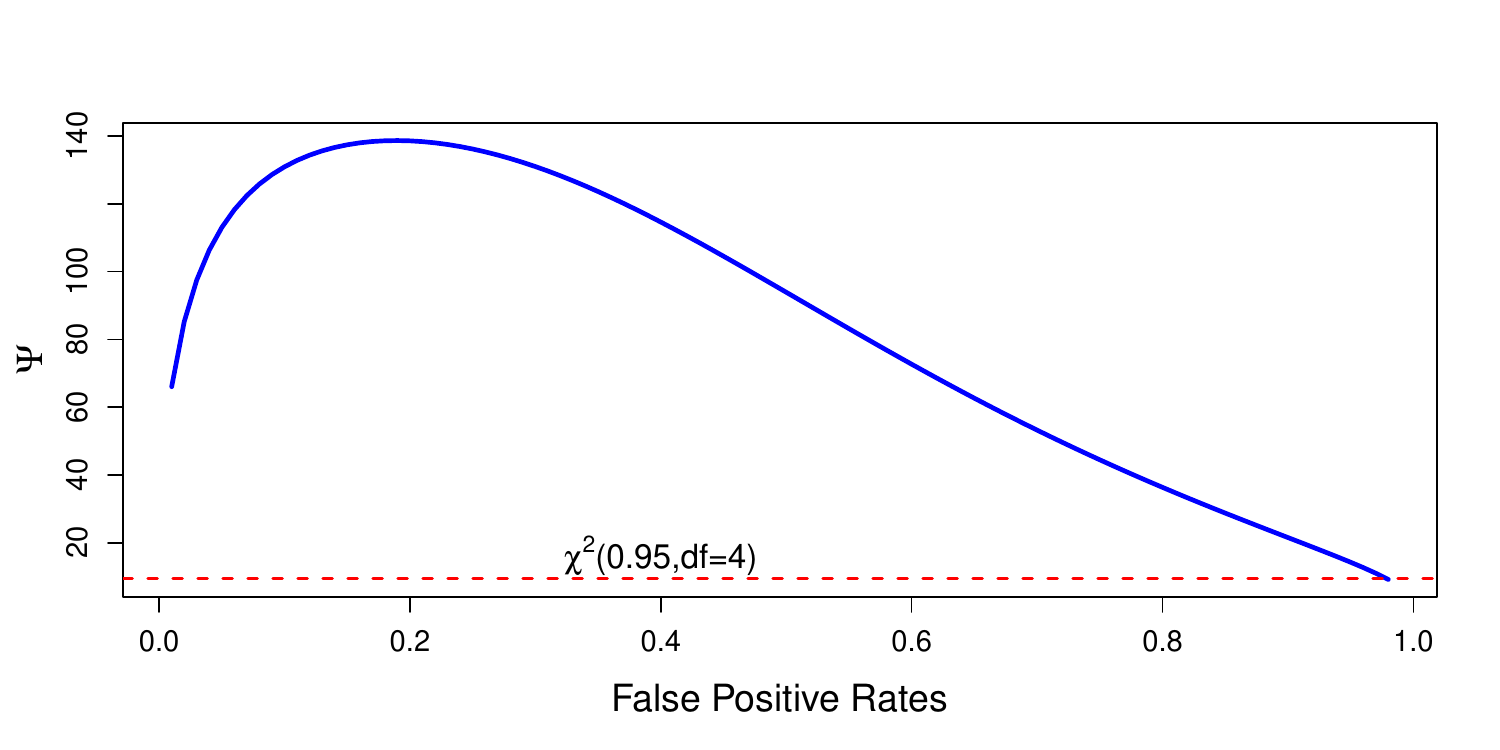}
\end{center}
\caption{ Plots the values of the test statistic $\Psi$ for ROC curves (the blue solid line) versus FPR. The dashed line is the critical value for the $95\%$ confidence of the Chi-square distribution with df=4.
 }
\label{fig:test_ROC}
\end{figure}

Next, we perform the homogeneity test for the ROCs, which  requires computing the test statistic $\Psi$ for all FPR values. In Figure \ref{fig:test_ROC} we plot the value of test $\Psi$  as a function of FPR. The test statistic is larger than the critical value $\chi^2\left(0.95,df=4\right)$, except for FPR values close to 1. Thus, the ROCs are different for FPRs smaller than 0.95 and the same for FPRs greater than 0.95. 

\subsection{Post hoc Pairwise Comparisons}

Since the homogeneity tests showed differences among the AUCs and ROCs of the five groups, we perform post hoc pairwise comparison. This allows us  to identify which groups have different AUCs or ROCs. First we compare AUCs, followed by ROCs. 

\begin{figure}
  \begin{center}
\includegraphics[height = 0.5\textheight]{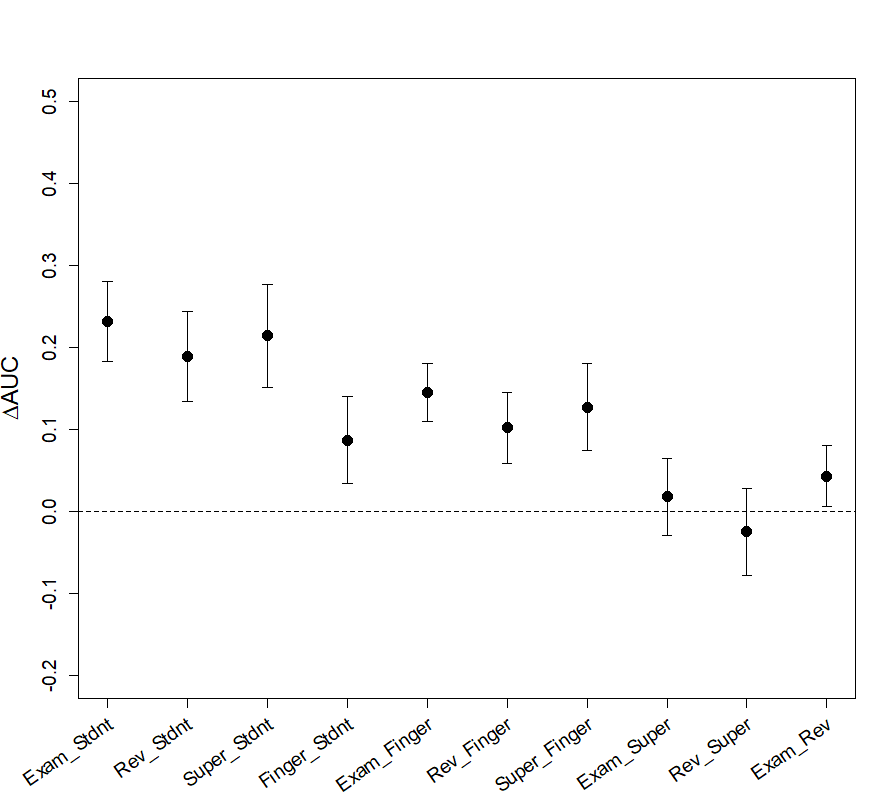}
\end{center}
\caption{Differences in AUCs among the five groups. The horizontal axis indicates the 10 pairwise comparisons between the groups. We shortened the group names, with Exam = examiners, Super = super-recognizers, Rev = reviewers, Finger = fingerprint examiners, and Stnt = students. The vertical axis reports the difference in AUC for the two groups in the comparison ($\Delta AUC$). The $\Delta AUC$ is given by the label in the horizontal axis, which is first minus second. For the Exam\_\_Stdnt, the difference is examiners' AUC minus students' AUC. For each comparison, we plot the estimated difference with a 95\% confidence interval. If the confidence interval is above the the zero line, then the difference is statistical significant. }
\label{fig:pair_AUC}
\end{figure}

For AUCs, we carried out pairwise comparisons between all five groups. Figure \ref{fig:pair_AUC} shows these comparisons. All our conclusions for pairwise comparison are with 95\% confidence. We concluded that two pairwise comparisons were the same:  facial examiners with super-recognizers and facial reviewers with super-recognizers. While our analysis found a difference between examiners and reviewers, \citet{Phillips:2018} did not. However, both analyses come to their conclusions by small margins. 

Next, we assess statistical differences between two ROCs, by comparing the ROCs at each FPR. For comparing two ROCs, there are three possible conclusions: the two ROCs are statistical the same for all FPR, they are statistical different for all FPR, or for some FPRs the two ROCs are the same and for some FPRs they are different. In our analysis we found all three cases.  In most applications, systems operate a low FPRs, and  our technique allows engineers to focus on the FPR relevant to their applications. 

\begin{figure}
  \begin{center}
\includegraphics[height = 0.52\textheight]{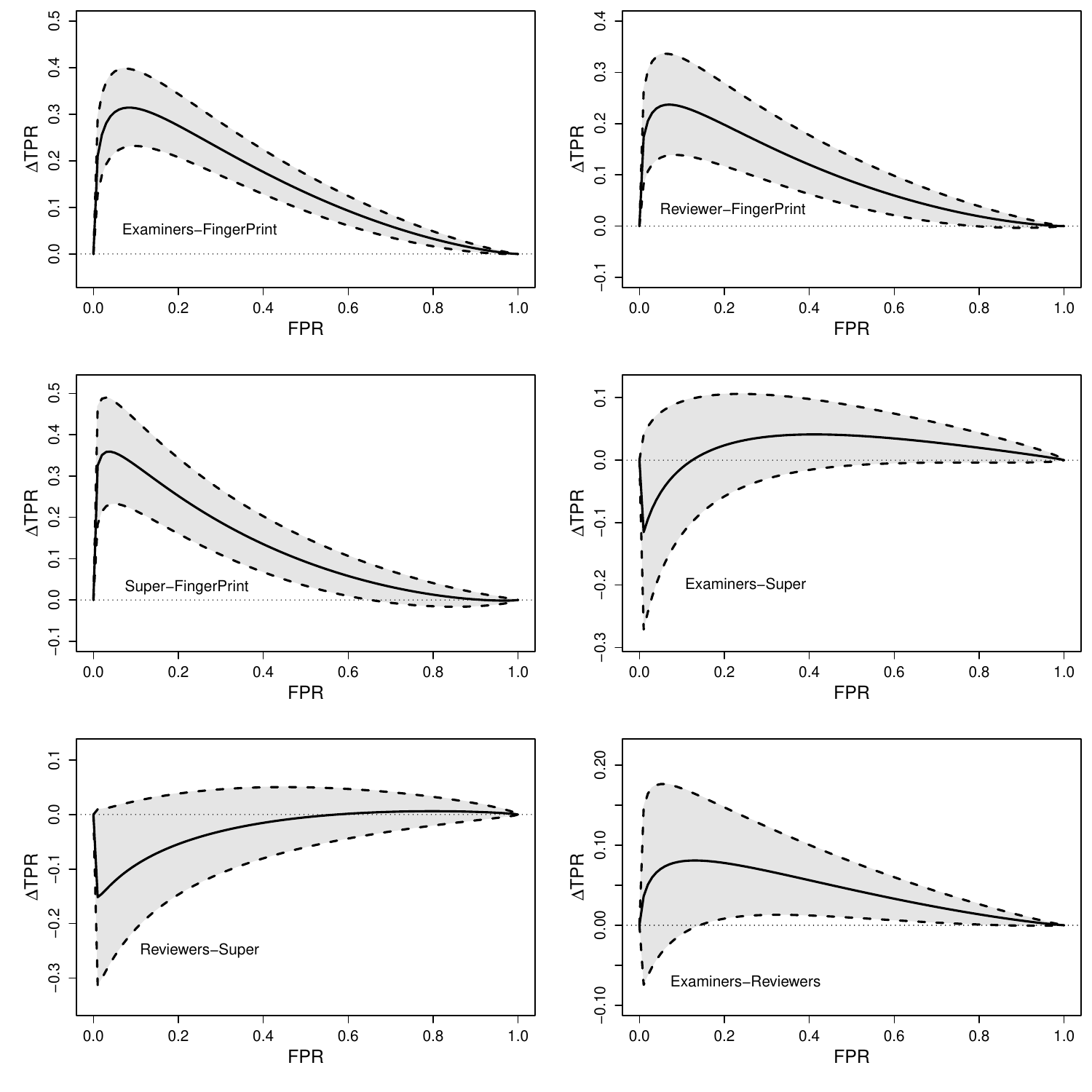}
\end{center}
\caption{Differences between ROC curves of four groups in facial recognition data. The horizontal axis corresponds to FPR. The vertical axis reports the $\Delta TPR$, the difference between the two ROCs at each FPR. The solid line shows the estimated difference between the ROCs.  Dashed lines are upper and lower bounds of the $95\%$ confidence band.
}
\label{fig:delROC}
\end{figure}

Figure \ref{fig:delROC} shows the pairwise comparison for four groups: examiners, reviewers, super-recognizers, and fingerprint examiners.  Comparisons with students can be found in the Appendix.

We start by looking at the pairwise comparison of facial examiners and fingerprint examiners, upper-left-hand plot in Figure \ref{fig:delROC}.  The horizontal axis corresponds to FPR, and the vertical axis reports the $\Delta TPR$, the difference between the two ROCs at each FPR. The solid line shows the estimated difference between the ROCs' for the face examiners minus the fingerprint examiners. Dashed lines are upper and lower bounds of the $95\%$ confidence band.  For all FPRs, the 95\% confidence band, the gray region, is above the $\Delta TPR = 0$ line. Thus, for the entire ROCs, the face examiners and fingerprint examiners are statistical different  with 95\% confidence.

For facial examiners and super-recognizers, the 95\% confidence band  contains the $\Delta TPR = 0$ line, therefore, the differences between the ROCs is not statistical significant with 95\% confidence for all FPRs. We get the same findings when comparing facial reviewer and super-recognizers. These results are consist with the previous ad-hoc analysis for AUCs that found no statistical difference with 95\% confidence. 

For facial examiners and reviewers, the confidence band is not above the $\Delta TPR = 0$ line, nor does the band contain the $\Delta TPR = 0$ line. Instead, for $FPR \le 0.15$ and $FPR > 0.8$, the the band contains the $\Delta TPR = 0$, and for $0.15 < FPR \le 0.8$, the the band contains the $\Delta TPR = 0$ line. Thus, for $FPR \le 0.15$ and $FPR > 0.8$, the examiners and reviewers have the same accuracy with a 95\% confidence, and for $0.15 < FPR \le 0.8$, the examiners and reviewers have different accuracy with a 95\% confidence.  In the majority of applications, the operating point requires a low FPR. Systems general operate at a low FPR to minimize false accusations. The comparison between super-recognizers and fingerprint examiners has a similar pattern. For $FPR < 0.6$, the difference is significant, and for $FPR > 0.6$, the difference is not significant--both with 95\% confidence. 


Overall, our conclusions are consistent with \citet{Phillips:2018} , with each having difference strengths.  \citet{Phillips:2018} concentrated on the accuracy of individual participants and permitted examination of the range of accuracy for members of each group. Our analysis treat groups as covariates, and analysis produced ROCs with confidence bands and AUCs with confidence intervals . One key strength of out approach is the ability to produce results at operationally relevant decision thresholds.  Since the majority of applications operate at low FPRs, producing results with error bands for ROCs will enable examiners, analysts and engineers to concentrate on the appropriate FPRs.  


\section{Conclusion}
\label{sec:conc}
In this paper, we have constructed a homogeneity test for covariate-adjusted ROC curves from ordinal scores from multiple rater groups. Estimators for ROC curves and the corresponding AUC were analytically formulated within the framework of ordinal regression combined with binormality assumption. Moreover, the asymptotic properties of the estimators were also computed. The simulation results have showed that at an appropriate sample size, the estimated ROC curve and corresponding AUC asymptotically converged. Then, we performed statistical inference based on estimated ROC curves and AUCs. Our study  indicates that confidence interval coverage of differences among ROC curves and among AUCs approached 95\% at a suitable sample size. Moreover, the fact that the type I error for the test based on ROC curves or on AUCs reached a rate of 5\% validated the test procedure.  Furthermore, for different settings that support alternative hypothesis, we determined the minimum sample sizes  numerically by taking advantage of their relationship to the power. Sample sizes were obtained at different levels of FPR. Simulations pointed out that the higher FPR required the larger sample size to achieve a desired power.   
Finally, we applied the test procedure to face recognition data consisting of five groups that included facial examiners, facial reviewers, super-recognizers, fingerprint examiners, and students. We found the identification accuracy highest for facial examiners and lowest for students. We concluded that differences in accuracy between facial examiners and super-recognizers, and facial reviewers and super-recognizers were statistically significant. The results for the post hoc, produced results that allow comparisons between groups at each FPR. This enables forensic examiners and engineers to focus on FPRs relevant to their applications.

\subsection*{Disclaimer} 
Opinions, recommendations, findings, and conclusions in this paper do not necessarily reflect the views or policies of NIST or the United States Government.

\subsection*{Acknowledgments}   
This research  was  supported in part by Award No. 2019-DU-BX-0011
awarded by the National Institute of Justice, Office of Justice Programs, US Department
of Justice. The opinions, fndings, and conclusions or recommendations expressed in this
publication are those of the authors and do not necessarily reflect those of the US Department
of Justice.
\bibliographystyle{apalike}
\bibliography{Ty_cite}

\bigskip
\section*{Appendix}
\subsection*{ Proof of Lemma 1}
\label{lemma}
The log likelihood $\ell_i \left(\boldsymbol \alpha,\boldsymbol \beta,\boldsymbol \tau \vert D_i,\textbf{x}_i,y_i\right)$ of the model \eqref{probit} for a single observation $D_i,\textbf{x}_i,y_i$ can be written as 
\begin{align*}
\ell_i \left(\boldsymbol \alpha,\boldsymbol \beta,\boldsymbol \tau \vert D_i,\textbf{x}_i,y_i\right) =\sum_{l=1}^{L}\left[y_{i}=l\right]\log \biggl[ \Phi\left\lbrace \frac{\tau_{l}-\left( \alpha_{0}D_i+\boldsymbol\alpha_{1}\textbf{x}_i+\boldsymbol\alpha_{2}D_i\textbf{x}_i\right)}{\exp\left(\beta_{0}D_i+\boldsymbol\beta_{1}\textbf{x}_i+\boldsymbol\beta_{2}D_i\textbf{x}_i\right)}\right\rbrace \\
-\Phi\left\lbrace \frac{\tau_{l-1}-\left( \alpha_{0}D_i+\boldsymbol\alpha_{1}\textbf{x}_i+\boldsymbol\alpha_{2}D_i\textbf{x}_i\right)}{\exp\left(\beta_{0}D_i+\boldsymbol\beta_{1}\textbf{x}_i+\boldsymbol\beta_{2}D_i\textbf{x}_i \right) }\right\rbrace  \biggl]
\end{align*}   
In above log likelihood function, $\boldsymbol{\alpha}$ denotes for $\alpha_0$, $\boldsymbol{\alpha}_1$ and $\boldsymbol{\alpha}_2$ and $\boldsymbol{\beta}$ does for $\beta_0$, $\boldsymbol{\beta}_1$ and $\boldsymbol{\beta}_2$.
\\Since $\lbrace \textbf{x}_i,y_i\rbrace_{i=1}^n$ are \textit{i.i.d} samples, the log likelihood function for the entire data (X,Y) is 
\begin{equation*}   
\ell_N\left(\boldsymbol \alpha,\boldsymbol \beta,\boldsymbol \tau\right)=\sum_{i=1}^{N}\ell_i\left(\boldsymbol \alpha,\boldsymbol \beta,\boldsymbol \tau \vert D_i,\textbf{x}_i,y_i\right) 
\end{equation*}

where $\textbf{x}_i$ is the $i^{th}$ row of X and $y_i$ is the $i^{th}$ element of vector Y.\\
The score function $\textbf{s}_N\left(\boldsymbol{\gamma}\right)\equiv \textbf{s}_N\left(\boldsymbol \alpha,\boldsymbol \beta,\boldsymbol \tau \right)$, which is the first order partial derivative of $\ell_N\left(\boldsymbol \alpha,\boldsymbol \beta,\boldsymbol \tau\right)$ with respect to $\boldsymbol \alpha,\boldsymbol \beta,\boldsymbol \tau$, is defined as 
\begin{equation*}
\textbf{s}_N\left(\boldsymbol{\gamma}\right)=\left(\textbf{s}_N\left(\boldsymbol \alpha\right),\textbf{s}_N\left(\boldsymbol \beta\right),\textbf{s}_N\left(\boldsymbol \tau\right)\right)
\end{equation*}
where 
$
 \textbf{s}_N\left(\boldsymbol{\alpha}\right)=\frac{\partial  \ell_N\left(\boldsymbol \alpha,\boldsymbol \beta,\boldsymbol \tau\right)}{\partial \boldsymbol{\alpha} }, \textbf{s}_N\left(\boldsymbol{\beta}\right)=\frac{\partial  \ell_N\left(\boldsymbol \alpha,\boldsymbol \beta,\boldsymbol \tau \right)}{\partial \boldsymbol{\beta} }, \textbf{s}_N\left(\boldsymbol{\tau}\right)=\frac{\partial  \ell_N\left(\boldsymbol \alpha,\boldsymbol \beta,\boldsymbol \tau \right)}{\partial \boldsymbol{\tau} }. $\\
 Here, $\boldsymbol{\gamma}$ is a composite notation of $\boldsymbol{\alpha}, \boldsymbol{\beta}, \boldsymbol{\tau} $.
The specific form of components of $\textbf{s}_N\left(\boldsymbol{\gamma}\right)$ is derived as
\\
$s_N\left(\alpha_0\right)=\sum_{i=1}^{N}\sum_{l=1}^{L}\left[y_i=l\right]\left(\frac{-D_i}{\xi_i}\right)\frac{\psi\left(\kappa_l\right)-\psi\left(\kappa_{l-1}\right)}{\left[\Phi\left(\kappa_l\right)-\Phi\left(\kappa_{l-1}\right)\right]}$
\\
$\textbf{s}_N\left(\boldsymbol{\alpha}_1\right)=\sum_{i=1}^{n}\sum_{l=1}^{L}\left[y_i=l\right]\left(\frac{-\textbf{x}_i}{\xi_i}\right)\frac{\psi\left(\kappa_l\right)-\psi\left(\kappa_{l-1}\right)}{\left[\Phi\left(\kappa_l\right)-\Phi\left(\kappa_{l-1}\right)\right]}$
\\
$\textbf{s}_N\left(\boldsymbol{\alpha}_2\right)=\sum_{i=1}^{n}\sum_{l=1}^{L}\left[y_i=l\right]\left(\frac{-D_i\textbf{x}_i}{\xi_i}\right)\frac{\psi\left(\kappa_l\right)-\psi\left(\kappa_{l-1}\right)}{\left[\Phi\left(\kappa_l\right)-\Phi\left(\kappa_{l-1}\right)\right]}$
\\
$s_N\left(\beta_0\right)=\sum_{i=1}^{n}\sum_{l=1}^{L}\left[y_i=l\right]\left(-D_i\right)\frac{\kappa_l\psi\left(\kappa_l\right)-\kappa_{l-1}\psi\left(\kappa_{l-1}\right)}{\Phi\left(\kappa_l\right)-\Phi\left(\kappa_{l-1}\right)}$
\\
$\textbf{s}_N\left(\boldsymbol{\beta}_1\right)=\sum_{i=1}^{n}\sum_{l=1}^{L}\left[y_i=l\right]\left(-\textbf{x}_i\right)\frac{\kappa_l\psi\left(\kappa_l\right)-\kappa_{l-1}\psi\left(\kappa_{l-1}\right)}{\Phi\left(\kappa_l\right)-\Phi\left(\kappa_{l-1}\right)}$
\\
$\textbf{s}_N\left(\boldsymbol{\beta}_2\right)=\sum_{i=1}^{n}\sum_{l=1}^{L}\left[y_i=l\right]\left(-D_i\textbf{x}_i\right)\frac{\kappa_l\psi\left(\kappa_l\right)-\kappa_{l-1}\psi\left(\kappa_{l-1}\right)}{\Phi\left(\kappa_l\right)-\Phi\left(\kappa_{l-1}\right)}$
\\
$s_N\left(\tau_l\right)=\sum_{i=1}^{n}\left[y_i=l\right]\frac{\psi\left(\kappa_l\right)}{\xi_i\left[\Phi\left(\kappa_l\right)-\Phi\left(\kappa_{l-1}\right)\right]}-\sum_{i=1}^{n}\left[y_i=l+1\right]\frac{\psi\left(\kappa_l\right)}{\xi_i\left[\Phi\left(\kappa_{l+1}\right)-\Phi\left(\kappa_{l}\right)\right]}$  with $1 \le l \le L-1$
where 
\\
$\xi_i= \exp\left(\beta_{0}D_i+\boldsymbol\beta_{1}\textbf{x}_i+\boldsymbol\beta_{2}D_i\textbf{x}_i\right) $ 
\\
$\kappa_l= \frac{\tau_{l}-\left( \alpha_{0}D_i+\boldsymbol\alpha_{1}\textbf{x}_i+\boldsymbol\alpha_{2}D_i\textbf{x}_i\right)}{\exp\left(\beta_{0}D_i+\boldsymbol\beta_{1}\textbf{x}_i+\boldsymbol\beta_{2}D_i\textbf{x}_i\right)} $ 
\\
$\psi\left(\kappa_l\right)=\frac{1}{\sqrt{2\pi}}e^{-\frac{1}{2}\kappa_l^2}$\\
Let $\boldsymbol{\gamma}_0$ and $\hat{\boldsymbol{\gamma}}$ be the true and the maximum likelihood estimated  vector of $\boldsymbol{\gamma}$.\\
Expand $\textbf{s}_N\left(\boldsymbol{\gamma}_0\right)$ about $\hat{\boldsymbol{\gamma}}$ and use the mean value theorem with the condition $\textbf{s}_N\left(\hat{\boldsymbol{\gamma}}\right)=\textbf{0}$, we obtain
\begin{equation}
\label{eq:mean_value}
\textbf{s}_N\left(\boldsymbol{\gamma}_0\right)=\left[ \int_0^1H_N\left(\boldsymbol{\gamma}_0+t\left(\hat{\boldsymbol{\gamma}}-\boldsymbol{\gamma}_0\right)\right)dt\right]\left(\hat{\boldsymbol{\gamma}}-\boldsymbol{\gamma}_0\right)
\end{equation}  
where 
$H_N\left(\boldsymbol{\gamma}\right)$, which is $(4p+L+1)\times (4p+L+1)$ matrix where $p$ is the length of $\textbf{x}_i$, is the first order partial derivative of $\textbf{s}_N\left(\boldsymbol{\gamma}\right)$ and in turn is the second order partial derivative  of $\ell_N\left(\boldsymbol{\gamma}\right)$ with respect to $\boldsymbol{\gamma}$. Due to the symmetric property of $H_N$, only diagonal elements and entries above the main diagonal are presented as follows.   \\
\begin{align*}
H_{1,1}=\frac{\partial s_N\left(\alpha_0\right)}{\partial \alpha_0}=\sum_{i=1}^{n}\sum_{l=1}^{L}\left[y_i=l\right] \frac{-D_i^2}{\xi_i^2}\left[\frac{\kappa_l\psi\left(\kappa_l\right) - \kappa_{l-1}\psi\left(\kappa_{l-1}\right)}{\Phi\left(\kappa_l\right)-\Phi\left(\kappa_{l-1}\right)}+ \frac{\left[\psi\left(\kappa_l\right)-\psi\left(\kappa_{l-1}\right)\right]^2}{\left[\Phi\left(\kappa_l\right)-\Phi\left(\kappa_{l-1}\right)\right]^2}\right]\\
H_{1,j+1}=\frac{\partial s_N\left(\alpha_0\right)}{\partial \alpha_{1j}}= \sum_{i=1}^{n}\sum_{l=1}^{L}\left[y_i=l\right] \frac{-D_i x_{ij}}{\xi_i^2}\left[\frac{\kappa_l\psi\left(\kappa_l\right) - \kappa_{l-1}\psi\left(\kappa_{l-1}\right)}{\Phi\left(\kappa_l\right)-\Phi\left(\kappa_{l-1}\right)}+ \frac{\left[\psi\left(\kappa_l\right)-\psi\left(\kappa_{l-1}\right)\right]^2}{\left[\Phi\left(\kappa_l\right)-\Phi\left(\kappa_{l-1}\right)\right]^2}\right]\\
 H_{1,p+1+j}=\frac{\partial s_N\left(\alpha_0\right)}{\partial \alpha_{2j}}= \sum_{i=1}^{n}\sum_{l=1}^{L}\left[y_i=l\right] \frac{-D_i^2 x_{ij}}{\xi_i^2}\left[\frac{\kappa_l^2\psi\left(\kappa_l\right) - \kappa_{l-1}\psi\left(\kappa_{l-1}\right)}{\Phi\left(\kappa_l\right)-\Phi\left(\kappa_{l-1}\right)}+ \frac{\left[\psi\left(\kappa_l\right)-\psi\left(\kappa_{l-1}\right)\right]^2}{\left[\Phi\left(\kappa_l\right)-\Phi\left(\kappa_{l-1}\right)\right]^2}\right]
\end{align*}
\begin{align*}
H_{1,2p+2}=\frac{\partial s_N\left(\alpha_0\right)}{\partial \beta_0}= \sum_{i=1}^{n}\sum_{l=1}^{L}\left[y_i=l\right] \frac{-D_i^2}{\xi_i}\biggl[\frac{\left(\kappa_l^2-1\right)\psi\left(\kappa_l\right)-\left(\kappa_{l-1}^2-1\right)\psi\left(\kappa_{l-1}\right)}{\Phi\left(\kappa_l\right)-\Phi\left(\kappa_{l-1}\right)}\\
+\frac{\left[\psi\left(\kappa_l\right)-\psi\left(\kappa_{l-1}\right)\right]\left[\kappa_l\psi\left(\kappa_l\right)-\kappa_{l-1}\psi\left(\kappa_{l-1}\right)\right]}{\left[\Phi\left(\kappa_l\right)-\Phi\left(\kappa_{l-1}\right)\right]^2}\biggl]
\end{align*}
\begin{align*}
H_{1,2p+2+j}=\frac{\partial s_N\left(\alpha_0\right)}{\partial \beta_{1j}}= \sum_{i=1}^{n}\sum_{l=1}^{L}\left[y_i=l\right] \frac{-D_i x_{ij}}{\xi_i}\biggl[\frac{\left(\kappa_l^2-1\right)\psi\left(\kappa_l\right) - \left(\kappa_{l-1}^2-1\right)\psi\left(\kappa_{l-1}\right)}{\Phi\left(\kappa_l\right)-\Phi\left(\kappa_{l-1}\right)}\\ +\frac{\left[\psi\left(\kappa_l\right)-\psi\left(\kappa_{l-1}\right)\right]\left[\kappa_l\psi\left(\kappa_l\right)-\kappa_{l-1}\psi\left(\kappa_{l-1}\right)\right]}{\left[\Phi\left(\kappa_l\right)-\Phi\left(\kappa_{l-1}\right)\right]^2}\biggl]
\end{align*}

\begin{align*}
H_{1,3p+2+j}=&\frac{\partial s_N\left(\alpha_0\right)}{\partial \beta_{2j}}= \sum_{i=1}^{n}\sum_{l=1}^{L}\left[y_i=l\right] \frac{-D_i^2x_{ij}}{\xi_i}\biggl[\frac{\left(\kappa_l^2-1\right)\psi\left(\kappa_l\right) - \left(\kappa_{l-1}^2-1\right)\psi\left(\kappa_{l-1}\right)}{\Phi\left(\kappa_l\right)-\Phi\left(\kappa_{l-1}\right)}\\
&+\frac{\left[\psi\left(\kappa_l\right)-\psi\left(\kappa_{l-1}\right)\right]\left[\kappa_l\psi\left(\kappa_l\right)-\kappa_{l-1}\psi\left(\kappa_{l-1}\right)\right]}{\left[\Phi\left(\kappa_l\right)-\Phi\left(\kappa_{l-1}\right)\right]^2}\biggl]
\end{align*}
\begin{align*}
 H_{1,4p+2+l}=&\frac{\partial s_N\left(\alpha_0\right)}{\partial \tau_{l}}= \sum_{i=1}^{n}\frac{D_i}{\xi_i^2} \left[y_i=l\right]\left(\frac{\kappa_l\psi\left(\kappa_l\right) }{\Phi\left(\kappa_l\right)-\Phi\left(\kappa_{l-1}\right)}+ \frac{\left[\psi\left(\kappa_l\right)-\psi\left(\kappa_{l-1}\right)\right]\psi\left(\kappa_l\right)}{\left[\Phi\left(\kappa_l\right)-\Phi\left(\kappa_{l-1}\right)\right]^2}\right)\\ 
 &-\sum_{i=1}^{n}\frac{D_i}{\xi_i^2} \left[y_i=l+1\right]\left(\frac{\kappa_l\psi\left(\kappa_l\right) }{\Phi\left(\kappa_{l+1}\right)-\Phi\left(\kappa_{l}\right)}+ \frac{\left[\psi\left(\kappa_{l+1}\right)-\psi\left(\kappa_{l}\right)\right]\psi\left(\kappa_{l}\right)}{\left[\Phi\left(\kappa_{l+1}\right)-\Phi\left(\kappa_{l}\right)\right]^2}\right) 
\end{align*}
\begin{align*}
H_{j+1,j_{1}+1}=\frac{\partial s_N\left(\alpha_{1j}\right)}{\partial \alpha_{1j_{1}}}= \sum_{i=1}^{n}\sum_{l=1}^{L}\left[y_i=l\right] \frac{-x_{ij}x_{ij_{1}}}{\xi_i^2}\left[\frac{\kappa_l\psi\left(\kappa_l\right) - \kappa_{l-1}\psi\left(\kappa_{l-1}\right)}{\Phi\left(\kappa_l\right)-\Phi\left(\kappa_{l-1}\right)}+ \frac{\left[\psi\left(\kappa_l\right)-\psi\left(\kappa_{l-1}\right)\right]^2}{\left[\Phi\left(\kappa_l\right)-\Phi\left(\kappa_{l-1}\right)\right]^2}\right]
\end{align*}

\begin{align*}H_{j+1,p+1+j_{2}}=\frac{\partial s_N\left(\alpha_{1j}\right)}{\partial \alpha_{2j_{2}}}= \sum_{i=1}^{n}\sum_{l=1}^{L}\left[y_i=l\right] \frac{-D_i x_{ij}x_{ij_{2}}}{\xi_i^2}\left[\frac{\kappa_l\psi\left(\kappa_l\right) - \kappa_{l-1}\psi\left(\kappa_{l-1}\right)}{\Phi\left(\kappa_l\right)-\Phi\left(\kappa_{l-1}\right)}+ \frac{\left[\psi\left(\kappa_l\right)-\psi\left(\kappa_{l-1}\right)\right]^2}{\left[\Phi\left(\kappa_l\right)-\Phi\left(\kappa_{l-1}\right)\right]^2}\right]
\end{align*}
\begin{align*}
H_{j+1,2p+2}=\frac{\partial s_N\left(\alpha_{1j}\right)}{\partial \beta_0}= \sum_{i=1}^{n}\sum_{l=1}^{L}\left[y_i=l\right] \frac{-D_ix_{ij}}{\xi_i}\biggl[\frac{\left(\kappa_l^2-1\right)\psi\left(\kappa_l\right)-\left(\kappa_{l-1}^2-1\right)\psi\left(\kappa_{l-1}\right)}{\Phi\left(\kappa_l\right)-\Phi\left(\kappa_{l-1}\right)}\\
+\frac{\left[\psi\left(\kappa_l\right)-\psi\left(\kappa_{l-1}\right)\right]\left[\kappa_l\psi\left(\kappa_l\right)-\kappa_{l-1}\psi\left(\kappa_{l-1}\right)\right]}{\left[\Phi\left(\kappa_l\right)-\Phi\left(\kappa_{l-1}\right)\right]^2}\biggl]
\end{align*}
\begin{align*}
H_{j+1,2p+2+j_2}=\frac{\partial s_N\left(\alpha_{1j}\right)}{\partial \beta_{1j_{2}}}= \sum_{i=1}^{n}\sum_{l=1}^{L}\left[y_i=l\right] \frac{-x_{ij}x_{ij_2}}{\xi_i}\biggl[\frac{\left(\kappa_l^2-1\right)\psi\left(\kappa_l\right)-\left(\kappa_{l-1}^2-1\right)\psi\left(\kappa_{l-1}\right)}{\Phi\left(\kappa_l\right)-\Phi\left(\kappa_{l-1}\right)}\\
+\frac{\left[\psi\left(\kappa_l\right)-\psi\left(\kappa_{l-1}\right)\right]\left[\kappa_l\psi\left(\kappa_l\right)-\kappa_{l-1}\psi\left(\kappa_{l-1}\right)\right]}{\left[\Phi\left(\kappa_l\right)-\Phi\left(\kappa_{l-1}\right)\right]^2}\biggl]
\end{align*}
\begin{align*}
H_{j+1,3p+2+j_2}=\frac{\partial s_N\left(\alpha_{1j}\right)}{\partial \beta_{2j_{2}}}= \sum_{i=1}^{n}\sum_{l=1}^{L}\left[y_i=l\right] \frac{-D_ix_{ij}x_{ij_2}}{\xi_i}\biggl[\frac{\left(\kappa_l^2-1\right)\psi\left(\kappa_l\right)-\left(\kappa_{l-1}^2-1\right)\psi\left(\kappa_{l-1}\right)}{\Phi\left(\kappa_l\right)-\Phi\left(\kappa_{l-1}\right)}\\
+\frac{\left[\psi\left(\kappa_l\right)-\psi\left(\kappa_{l-1}\right)\right]\left[\kappa_l\psi\left(\kappa_l\right)-\kappa_{l-1}\psi\left(\kappa_{l-1}\right)\right]}{\left[\Phi\left(\kappa_l\right)-\Phi\left(\kappa_{l-1}\right)\right]^2}\biggl]
\end{align*}
\begin{align*}
H_{j+1,4p+2+l}=\frac{\partial s_N\left(\alpha_{1j}\right)}{\partial \tau_{l}}= \sum_{i=1}^{n}\frac{ x_{ij}}{\xi_i^2} \left[y_i=l\right]\left(\frac{\kappa_l\psi\left(\kappa_l\right) }{\Phi\left(\kappa_l\right)-\Phi\left(\kappa_{l-1}\right)}+ \frac{\left[\psi\left(\kappa_l\right)-\psi\left(\kappa_{l-1}\right)\right]\psi\left(\kappa_l\right)}{\left[\Phi\left(\kappa_l\right)-\Phi\left(\kappa_{l-1}\right)\right]^2}\right)\\ 
 -\sum_{i=1}^{n}\frac{x_{ij}}{\xi_i^2} \left[y_i=l+1\right]\left(\frac{\kappa_l\psi\left(\kappa_l\right) }{\Phi\left(\kappa_{l+1}\right)-\Phi\left(\kappa_{l}\right)}+ \frac{\left[\psi\left(\kappa_{l+1}\right)-\psi\left(\kappa_{l}\right)\right]\psi\left(\kappa_{l}\right)}{\left[\Phi\left(\kappa_{l+1}\right)-\Phi\left(\kappa_{l}\right)\right]^2}\right) 
\end{align*}
\begin{align*}
H_{p+1+j,p+1+j_1}=\frac{\partial s_N\left(\alpha_{2j}\right)}{\partial \alpha_{2j_1}}= \sum_{i=1}^{n}\sum_{l=1}^{L}\left[y_i=l\right] \frac{-D_i^2 x_{ij}x_{ij_{1}}}{\xi_i^2}\left[\frac{\kappa_l\psi\left(\kappa_l\right) - \kappa_{l-1}\psi\left(\kappa_{l-1}\right)}{\Phi\left(\kappa_l\right)-\Phi\left(\kappa_{l-1}\right)}+ \frac{\left[\psi\left(\kappa_l\right)-\psi\left(\kappa_{l-1}\right)\right]^2}{\left[\Phi\left(\kappa_l\right)-\Phi\left(\kappa_{l-1}\right)\right]^2}\right]
\end{align*}
\begin{align*}
H_{p+1+j,2p+2}=\frac{\partial s_N\left(\alpha_{2j}\right)}{\partial \beta_0}= \sum_{i=1}^{n}\sum_{l=1}^{L}\left[y_i=l\right] \frac{-D_i^2x_{ij}}{\xi_i}\biggl[\frac{\left(\kappa_l^2-1\right)\psi\left(\kappa_l\right)-\left(\kappa_{l-1}^2-1\right)\psi\left(\kappa_{l-1}\right)}{\Phi\left(\kappa_l\right)-\Phi\left(\kappa_{l-1}\right)}\\
+\frac{\left[\psi\left(\kappa_l\right)-\psi\left(\kappa_{l-1}\right)\right]\left[\kappa_l\psi\left(\kappa_l\right)-\kappa_{l-1}\psi\left(\kappa_{l-1}\right)\right]}{\left[\Phi\left(\kappa_l\right)-\Phi\left(\kappa_{l-1}\right)\right]^2}\biggl]
\end{align*} 
\begin{align*}
H_{p+1+j,2p+2+j_2}=\frac{\partial s_N\left(\alpha_{2j}\right)}{\partial \beta_{1j_2}}= \sum_{i=1}^{n}\sum_{l=1}^{L}\left[y_i=l\right] \frac{-D_i x_{ij}x_{ij_2}}{\xi_i}\biggl[\frac{\left(\kappa_l^2-1\right)\psi\left(\kappa_l\right)-\left(\kappa_{l-1}^2-1\right)\psi\left(\kappa_{l-1}\right)}{\Phi\left(\kappa_l\right)-\Phi\left(\kappa_{l-1}\right)}\\
+\frac{\left[\psi\left(\kappa_l\right)-\psi\left(\kappa_{l-1}\right)\right]\left[\kappa_l\psi\left(\kappa_l\right)-\kappa_{l-1}\psi\left(\kappa_{l-1}\right)\right]}{\left[\Phi\left(\kappa_l\right)-\Phi\left(\kappa_{l-1}\right)\right]^2}\biggl]
\end{align*} 
\begin{align*}
H_{p+1+j,3p+2+j_2}=\frac{\partial s_N\left(\alpha_{2j}\right)}{\partial \beta_{2j_2}}= \sum_{i=1}^{n}\sum_{l=1}^{L}\left[y_i=l\right] \frac{-D_i^2x_{ij}x_{ij_2}}{\xi_i}\biggl[\frac{\left(\kappa_l^2-1\right)\psi\left(\kappa_l\right)-\left(\kappa_{l-1}^2-1\right)\psi\left(\kappa_{l-1}\right)}{\Phi\left(\kappa_l\right)-\Phi\left(\kappa_{l-1}\right)}\\
+\frac{\left[\psi\left(\kappa_l\right)-\psi\left(\kappa_{l-1}\right)\right]\left[\kappa_l\psi\left(\kappa_l\right)-\kappa_{l-1}\psi\left(\kappa_{l-1}\right)\right]}{\left[\Phi\left(\kappa_l\right)-\Phi\left(\kappa_{l-1}\right)\right]^2}\biggl]
\end{align*} 
\begin{align*}
H_{p+1+j,4p+2+l}=\frac{\partial s_N\left(\alpha_{2j}\right)}{\partial \tau_{l}}=\sum_{i=1}^{n}\frac{D_ix_{ij}}{\xi_i^2} \left[y_i=l\right]\left(\frac{\kappa_l\psi\left(\kappa_l\right) }{\Phi\left(\kappa_l\right)-\Phi\left(\kappa_{l-1}\right)}+ \frac{\left[\psi\left(\kappa_l\right)-\psi\left(\kappa_{l-1}\right)\right]\psi\left(\kappa_l\right)}{\left[\Phi\left(\kappa_l\right)-\Phi\left(\kappa_{l-1}\right)\right]^2}\right)\\ 
 -\sum_{i=1}^{n}\frac{D_i x_{ij}}{\xi_i^2} \left[y_i=l+1\right]\left(\frac{\kappa_l\psi\left(\kappa_l\right) }{\Phi\left(\kappa_{l+1}\right)-\Phi\left(\kappa_{l}\right)}+ \frac{\left[\psi\left(\kappa_{l+1}\right)-\psi\left(\kappa_{l}\right)\right]\psi\left(\kappa_{l}\right)}{\left[\Phi\left(\kappa_{l+1}\right)-\Phi\left(\kappa_{l}\right)\right]^2}\right) 
\end{align*} 
\begin{align*}
H_{2p+1,2p+1}=\frac{\partial s_N\left(\beta_{0}\right)}{\partial \beta_{0}}=\sum_{i=1}^{n}\sum_{l=1}^{L}\left[y_i=l\right]D_i^2\biggl[\frac{\kappa_l\left(1-\kappa_l^2\right)\psi\left(\kappa_l\right)-\kappa_{l-1}\left(1-\kappa_{l-1}^2\right)\psi\left(\kappa_{l-1}\right)}{\Phi\left(\kappa_l\right)-\Phi\left(\kappa_{l-1}\right)}\\
-\frac{\left[\kappa_l\psi\left(\kappa_l\right)-\kappa_{l-1}\psi\left(\kappa_{l-1}\right)\right]^2}{\left[\Phi\left(\kappa_l\right)-\Phi\left(\kappa_{l-1}\right)\right]^2}\biggl]
\end{align*} 
\begin{align*}
H_{2p+1,2p+1+j_2}=\frac{\partial s_N\left(\beta_{0}\right)}{\partial \beta_{1j_2}}=\sum_{i=1}^{n}\sum_{l=1}^{L}\left[y_i=l\right]D_i x_{ij_2}\biggl[\frac{\kappa_l\left(1-\kappa_l^2\right)\psi\left(\kappa_l\right)-\kappa_{l-1}\left(1-\kappa_{l-1}^2\right)\psi\left(\kappa_{l-1}\right)}{\Phi\left(\kappa_l\right)-\Phi\left(\kappa_{l-1}\right)}\\
-\frac{\left[\kappa_l\psi\left(\kappa_l\right)-\kappa_{l-1}\psi\left(\kappa_{l-1}\right)\right]^2}{\left[\Phi\left(\kappa_l\right)-\Phi\left(\kappa_{l-1}\right)\right]^2}\biggl]
\end{align*} 
\begin{align*}
H_{2p+1,3p+1+j_2}=\frac{\partial s_N\left(\beta_{0}\right)}{\partial \beta_{2j_2}}=\sum_{i=1}^{n}\sum_{l=1}^{L}\left[y_i=l\right]D_i^2 x_{ij_2}\biggl[\frac{\kappa_l\left(1-\kappa_l^2\right)\psi\left(\kappa_l\right)-\kappa_{l-1}\left(1-\kappa_{l-1}^2\right)\psi\left(\kappa_{l-1}\right)}{\Phi\left(\kappa_l\right)-\Phi\left(\kappa_{l-1}\right)}\\
-\frac{\left[\kappa_l\psi\left(\kappa_l\right)-\kappa_{l-1}\psi\left(\kappa_{l-1}\right)\right]^2}{\left[\Phi\left(\kappa_l\right)-\Phi\left(\kappa_{l-1}\right)\right]^2}\biggl]
\end{align*} 
\begin{align*}
H_{2p+1,4p+1+l}=\frac{\partial s_N\left(\beta_{0}\right)}{\partial \tau_{l}}=\sum_{i=1}^{n}\sum_{l=1}^{L}\left[y_i=l\right] \frac{D_i}{\xi_i}\biggl[\frac{\left(\kappa_l^2-1\right)\psi\left(\kappa_l\right)-\left(\kappa_{l-1}^2-1\right)\psi\left(\kappa_{l-1}\right)}{\Phi\left(\kappa_l\right)-\Phi\left(\kappa_{l-1}\right)}\\
+\frac{\left[\psi\left(\kappa_l\right)-\psi\left(\kappa_{l-1}\right)\right]\left[\kappa_l\psi\left(\kappa_l\right)-\kappa_{l-1}\psi\left(\kappa_{l-1}\right)\right]}{\left[\Phi\left(\kappa_l\right)-\Phi\left(\kappa_{l-1}\right)\right]^2}\biggl]
\end{align*}
\begin{align*}
H_{2p+1+j,2p+1+j_1}=\frac{\partial s_N\left(\beta_{1j}\right)}{\partial \beta_{1j_1}}=\sum_{i=1}^{n}\sum_{l=1}^{L}\left[y_i=l\right]x_{ij} x_{ij_1}\biggl[\frac{\kappa_l\left(1-\kappa_l^2\right)\psi\left(\kappa_l\right)-\kappa_{l-1}\left(1-\kappa_{l-1}^2\right)\psi\left(\kappa_{l-1}\right)}{\Phi\left(\kappa_l\right)-\Phi\left(\kappa_{l-1}\right)}\\
-\frac{\left[\kappa_l\psi\left(\kappa_l\right)-\kappa_{l-1}\psi\left(\kappa_{l-1}\right)\right]^2}{\left[\Phi\left(\kappa_l\right)-\Phi\left(\kappa_{l-1}\right)\right]^2}\biggl]
\end{align*}
\begin{align*}
H_{2p+1+j,3p+1+j_2}=\frac{\partial s_N\left(\beta_{1j}\right)}{\partial \beta_{2j_2}}=\sum_{i=1}^{n}\sum_{l=1}^{L}\left[y_i=l\right]D_ix_{ij} x_{ij_2}\biggl[\frac{\kappa_l\left(1-\kappa_l^2\right)\psi\left(\kappa_l\right)-\kappa_{l-1}\left(1-\kappa_{l-1}^2\right)\psi\left(\kappa_{l-1}\right)}{\Phi\left(\kappa_l\right)-\Phi\left(\kappa_{l-1}\right)}\\
-\frac{\left[\kappa_l\psi\left(\kappa_l\right)-\kappa_{l-1}\psi\left(\kappa_{l-1}\right)\right]^2}{\left[\Phi\left(\kappa_l\right)-\Phi\left(\kappa_{l-1}\right)\right]^2}\biggl]
\end{align*}
\begin{align*}
H_{2p+1+j,4p+1+l}=\frac{\partial s_N\left(\beta_{1j}\right)}{\partial \tau_{l}}=\sum_{i=1}^{n}\sum_{l=1}^{L}\left[y_i=l\right] \frac{x_{ij}}{\xi_i}\biggl[\frac{\left(\kappa_l^2-1\right)\psi\left(\kappa_l\right)-\left(\kappa_{l-1}^2-1\right)\psi\left(\kappa_{l-1}\right)}{\Phi\left(\kappa_l\right)-\Phi\left(\kappa_{l-1}\right)}\\
+\frac{\left[\psi\left(\kappa_l\right)-\psi\left(\kappa_{l-1}\right)\right]\left[\kappa_l\psi\left(\kappa_l\right)-\kappa_{l-1}\psi\left(\kappa_{l-1}\right)\right]}{\left[\Phi\left(\kappa_l\right)-\Phi\left(\kappa_{l-1}\right)\right]^2}\biggl]
\end{align*}
\begin{align*}
H_{3p+1+j,3p+1+j_2}=\frac{\partial s_N\left(\beta_{2j}\right)}{\partial \beta_{2j_2}}=\sum_{i=1}^{n}\sum_{l=1}^{L}\left[y_i=l\right]D_i^2x_{ij} x_{ij_2}\biggl[\frac{\kappa_l\left(1-\kappa_l^2\right)\psi\left(\kappa_l\right)-\kappa_{l-1}\left(1-\kappa_{l-1}^2\right)\psi\left(\kappa_{l-1}\right)}{\Phi\left(\kappa_l\right)-\Phi\left(\kappa_{l-1}\right)}\\
-\frac{\left[\kappa_l\psi\left(\kappa_l\right)-\kappa_{l-1}\psi\left(\kappa_{l-1}\right)\right]^2}{\left[\Phi\left(\kappa_l\right)-\Phi\left(\kappa_{l-1}\right)\right]^2}\biggl]
\end{align*}
\begin{align*}
H_{3p+1+j,4p+1+l}=\frac{\partial s_N\left(\beta_{2j}\right)}{\partial \tau_{l}}=\sum_{i=1}^{n}\sum_{l=1}^{L}\left[y_i=l\right] \frac{D_ix_{ij}}{\xi_i}\biggl[\frac{\left(\kappa_l^2-1\right)\psi\left(\kappa_l\right)-\left(\kappa_{l-1}^2-1\right)\psi\left(\kappa_{l-1}\right)}{\Phi\left(\kappa_l\right)-\Phi\left(\kappa_{l-1}\right)}\\
+\frac{\left[\psi\left(\kappa_l\right)-\psi\left(\kappa_{l-1}\right)\right]\left[\kappa_l\psi\left(\kappa_l\right)-\kappa_{l-1}\psi\left(\kappa_{l-1}\right)\right]}{\left[\Phi\left(\kappa_l\right)-\Phi\left(\kappa_{l-1}\right)\right]^2}\biggl]
\end{align*}
\begin{align*}
H_{4p+1+l,4p+1+l}=\frac{\partial s_N\left(\tau_{l}\right)}{\partial \tau_{l}}=
\sum_{i=1}^{n}\left[y_i=l\right] \left(\frac{-1}{\xi_i^2}\right)\biggl[\frac{\kappa_l\psi\left(\kappa_l\right)}{\Phi\left(\kappa_l\right)-\Phi\left(\kappa_{l-1}\right)}+\frac{\psi^2\left(\kappa_l\right)}{\left[\Phi\left(\kappa_l\right)-\Phi\left(\kappa_{l-1}\right)\right]^2}\biggl]\\
+\sum_{i=1}^{n}\left[y_i=l+1\right] \left(\frac{1}{\xi_i^2}\right)\biggl[\frac{\kappa_l\psi\left(\kappa_l\right)}{\Phi\left(\kappa_{l+1}\right)-\Phi\left(\kappa_{l}\right)}-\frac{\psi^2\left(\kappa_l\right)}{\left[\Phi\left(\kappa_{l+1}\right)-\Phi\left(\kappa_{l}\right)\right]^2}\biggl]
\end{align*}
\begin{align*}
H_{4p+1+l,4p+1+l+1}=\frac{\partial s_N\left(\tau_{l}\right)}{\partial \tau_{l+1}}=
\sum_{i=1}^{n}\left[y_i=l+1\right] \left(\frac{1}{\xi_i^2}\right)\frac{\psi\left(\kappa_l\right)\psi\left(\kappa_{l+1}\right)}{\left[\Phi\left(\kappa_{l+1}\right)-\Phi\left(\kappa_{l}\right)\right]^2}
\end{align*}
\begin{align*}
H_{4p+1+l,4p+1+l^{\prime}}=\frac{\partial s_N\left(\tau_{l}\right)}{\partial \tau_{l^{\prime}}}=0
\end{align*}
with $1 \le j \le p$, $j \le j_{1} \le p, $
  $1 \le j_{2} \le p$, $1 \le l \le L-1$, $l+1 < l^{\prime} \le L-1$.
\\From equation \eqref{eq:mean_value}, 
\begin{multline}
\label{eq:cond}
H_N^{-1/2}\left(\boldsymbol{\gamma}_0\right) \textbf{s}_N\left(\boldsymbol{\gamma}_0\right)=\left[ \int_0^1 H_N^{-1/2}\left(\boldsymbol{\gamma}_0\right)H_N\left(\boldsymbol{\gamma}_0+t\left(\hat{\boldsymbol{\gamma}}-\boldsymbol{\gamma}_0\right)\right)H_N^{-T/2}\left(\boldsymbol{\gamma}_0\right)dt\right] H_N^{T/2}\left(\boldsymbol{\gamma}_0\right)\left(\hat{\boldsymbol{\gamma}}-\boldsymbol{\gamma}_0\right)\\
=\left[ \int_0^1 G_N\left(\boldsymbol{\gamma}_0+t\left(\hat{\boldsymbol{\gamma}}-\boldsymbol{\gamma}_0\right)\right)dt\right] H_N^{T/2}\left(\boldsymbol{\gamma}_0\right)\left(\hat{\boldsymbol{\gamma}}-\boldsymbol{\gamma}_0\right)
\end{multline} 
where $G_N\left(\boldsymbol{\gamma}_0+t\left(\hat{\boldsymbol{\gamma}}-\boldsymbol{\gamma}_0\right)\right)=H_N^{-1/2}\left(\boldsymbol{\gamma}_0\right)H_N\left(\boldsymbol{\gamma}+t\left(\hat{\boldsymbol{\gamma}}-\boldsymbol{\gamma}_0\right)\right)H_N^{-T/2}\left(\boldsymbol{\gamma}_0\right)$\\
To prove $ H_N^{T/2}\left(\boldsymbol{\gamma}_0\right)\left(\hat{\boldsymbol{\gamma}}-\boldsymbol{\gamma}_0\right) \xrightarrow[]{} N\left(\textbf{0},I\right)$, I will show $\left[ \int_0^1 G_N\left(\boldsymbol{\gamma}_0+t\left(\hat{\boldsymbol{\gamma}}-\boldsymbol{\gamma}_0\right)\right)dt\right] \xrightarrow[]{} I$ and $H_N^{-1/2}\left(\boldsymbol{\gamma}_0\right) \textbf{s}_N\left(\boldsymbol{\gamma}_0\right) \xrightarrow[]{} N\left(\textbf{0},I\right)$
\begin{itemize}
    \item Prove $\left[ \int_0^1
    G_N\left(\boldsymbol{\gamma}_0+t\left(\hat{\boldsymbol{\gamma}}-\boldsymbol{\gamma}_0\right)\right)dt\right] \xrightarrow[]{} I$
\end{itemize}
Define the sequence $\aleph_N\left(\delta\right)$, $\delta > 0$, of neighborhoods of $\boldsymbol{\gamma}_0$ as\\
$\aleph_N\left(\delta\right)=\lbrace \boldsymbol{\gamma}: \left\Vert H_N^{T/2}\left(\boldsymbol{\gamma}-\boldsymbol{\gamma}_0\right)\le \delta \right\Vert \rbrace,\quad n=1,2,\cdots$\\
The following assumption needs to be satisfied.\\ That is, for all $\delta$, 
$
    \max_{\boldsymbol{\gamma} \in \aleph_N\left(\delta\right) }\left\Vert G_N\left(\boldsymbol{\gamma}\right)-I\right\Vert \rightarrow 0.
$
Using this assumption and $\left\Vert \int \cdot \right\Vert \leq \int \left\Vert \cdot \right\Vert$, we have, for any $\epsilon > 0$ 
\begin{equation*}
   \left\Vert \int_0^1G_N\left(\boldsymbol{\gamma}_0+t\left(\hat{\boldsymbol{\gamma}}-\boldsymbol{\gamma}_0\right)\right)dt - \textbf{I} \right\Vert \leq \int_0^1 \epsilon dt=\epsilon 
\end{equation*}
if $n$ is large and if, with some $\delta > 0$, $\boldsymbol{\hat{\gamma}}$ is in $\aleph_N\left(\delta\right)$, $\delta$ can be chosen so that the probability of this event is close to 1. Thus, 
\begin{equation}
\label{eq:cond1}
    \int_0^1G_N\left(\boldsymbol{\gamma}_0+t\left(\hat{\boldsymbol{\gamma}}-\boldsymbol{\gamma}_0\right)\right)dt \xrightarrow[]{p} \textbf{I} 
\end{equation}
\begin{itemize}
    \item Next, we can prove that $H_N^{-1/2}\left(\boldsymbol{\gamma}_0\right) \textbf{s}_N\left(\boldsymbol{\gamma}_0\right) \xrightarrow[]{d} N\left(\textbf{0},\textbf{I}\right)$
\end{itemize}
We prove the above convergence by indicating that the moment generating function (\textit{mgf})
$E\left[\delta \boldsymbol{\lambda}^{\prime}H_N^{-1/2}\left(\boldsymbol{\gamma}_0\right) \textbf{s}_N\left(\boldsymbol{\gamma}_0\right)\right]$
with $\boldsymbol{\lambda}^{\prime}\boldsymbol{\lambda}=1$ converges to that of the standard normal distribution.\\
For the sequence, $\boldsymbol{\gamma}_N=\boldsymbol{\gamma}_0+\delta H_N^{-T/2}\left(\boldsymbol{\gamma}_0\right)\boldsymbol{\lambda}, n=1,2,\cdots$, we have $\boldsymbol{\gamma}_N \in \aleph_N\left(\delta\right).$
Apply Taylor expansion for $\ell_N\left(\boldsymbol{\gamma}_N\right)$, we obtain 
\begin{equation}
\label{eq:Taylor}
\ell_N\left(\boldsymbol{\gamma}_N\right)=\ell_N\left(\boldsymbol{\gamma}_0\right)+\left(\boldsymbol{\gamma}_N-\boldsymbol{\gamma}_0\right)^{\prime}\textbf{s}_N\left(\boldsymbol{\gamma}_0\right)-\left(\boldsymbol{\gamma}_N-\boldsymbol{\gamma}_0\right)^{\prime}H_N\left(\Tilde{\gamma}_N\right)\left(\boldsymbol{\gamma}_N-\boldsymbol{\gamma}_0\right)/2    
\end{equation}
where $\Tilde{\gamma}_N$ on the line segment between $\boldsymbol{\gamma}_N$ and $\boldsymbol{\gamma}_0$. Substitute $\boldsymbol{\gamma}_N-\boldsymbol{\gamma}_0=\delta H_N^{-T/2}\left(\boldsymbol{\gamma}_0\right)\boldsymbol{\lambda}$ into \eqref{eq:Taylor} yields to 
\begin{equation*}
\label{eq:Taylor2}
\ell_N\left(\boldsymbol{\gamma}_N\right)=\ell_N\left(\boldsymbol{\gamma}_0\right)+\delta\boldsymbol{\lambda}^{\prime}H_N^{-1/2}\left(\boldsymbol{\gamma}_0\right)\textbf{s}_N\left(\boldsymbol{\gamma}_0\right)-\frac{\delta^2}{2}\boldsymbol{\lambda}^{\prime}H_N^{-1/2}\left(\boldsymbol{\gamma}_0\right)H_N\left(\Tilde{\gamma}_N\right)H_N^{-T/2}\left(\boldsymbol{\gamma}_0\right)\boldsymbol{\lambda}
\end{equation*}
or 
\begin{equation*}
\ell_N\left(\boldsymbol{\gamma}_N\right)=\ell_N\left(\boldsymbol{\gamma}_0\right)+\delta\boldsymbol{\lambda}^{\prime}H_N^{-1/2}\left(\boldsymbol{\gamma}_0\right)\textbf{s}_N\left(\boldsymbol{\gamma}_0\right)-\frac{\delta^2}{2}\boldsymbol{\lambda}^{\prime}G_N\left(\Tilde{\gamma}_N\right)\boldsymbol{\lambda}
\end{equation*}
Taking exponentials above equation and repositioning leads to 
\begin{equation*}
    \exp\left(\boldsymbol{\lambda}^{\prime}G_{n}\left( \Tilde{\gamma}_N\right)\boldsymbol{\lambda}\delta^2/2\right)L_N\left(\boldsymbol{\gamma}_N\right)=\exp\left(\delta\boldsymbol{\lambda}^{\prime}H_N^{-1/2}\left(\boldsymbol{\gamma}_0\right)\textbf{s}_N\left(\boldsymbol{\gamma}_0\right)\right)L_N\left(\boldsymbol{\gamma}_0\right)
\end{equation*}
where $L_N\left(\boldsymbol{\gamma}_N\right)$ is the likelihood.
Integrating both sides, we get 
\begin{equation*}
    \textbf{E}_{\boldsymbol{\gamma}_N}\exp\left(\boldsymbol{\lambda}^{\prime}G_{n}\left( \Tilde{\gamma}_N\right)\boldsymbol{\lambda}\delta^2/2\right)=\textbf{E}\exp\left(\delta\boldsymbol{\lambda}^{\prime}H_N^{-1/2}\left(\boldsymbol{\gamma}_0\right)\textbf{s}_N\left(\boldsymbol{\gamma}_0\right)\right)
\end{equation*}
Using conditions $
    \max_{\Tilde{\boldsymbol{\gamma}}_N \in \aleph_N\left(\delta\right) }\left\Vert G_N\left(\Tilde{\boldsymbol{\gamma}}_N\right)-I\right\Vert \rightarrow 0
$ and $\boldsymbol{\lambda}^{\prime}\boldsymbol{\lambda}=1$ we have, for any $\epsilon >0$, there exists a number $n_1$ with 
\begin{equation*}
\left\vert \exp\left(\boldsymbol{\lambda}^{\prime}G_{n}\left( \Tilde{\gamma}_N\right)\boldsymbol{\lambda}\delta^2/2\right)-\exp\left(\delta^2/2\right) \le \epsilon\right\vert ,\qquad n\ge n_1   
\end{equation*}
Integration of the above inequality with $\left\vert\int \cdot\right\vert \le \int \left\vert \cdot \right\vert$ indicates that 
$\textbf{E}_{\boldsymbol{\gamma}_N}\exp\left(\boldsymbol{\lambda}^{\prime}G_{n}\left( \Tilde{\gamma}_N\right)\boldsymbol{\lambda}\delta^2/2\right)$ and then $\textbf{E}\exp\left(\delta\boldsymbol{\lambda}^{\prime}H_{n}^{-1/2}\left(\boldsymbol{\gamma}_0\right)\textbf{s}_N\left(\boldsymbol{\gamma}_0\right)\right)$ converges to $\exp\left(\delta^2/2\right)$ which is the $mgf$ of the standard normal distribution. Because $\boldsymbol{\lambda}$ is an arbitrary unit vector, we eventually have 
\begin{equation}
\label{eq:cond2}
H_{n}^{-1/2}\left(\boldsymbol{\gamma}_0\right)\textbf{s}_N\left(\boldsymbol{\gamma}_0\right) \xrightarrow[]{d} N\left(\textbf{0},\textbf{I}\right)  
\end{equation}
\\From \eqref{eq:cond},\eqref{eq:cond1},\eqref{eq:cond2}, we obtain 
$H_N^{T/2}\left(\boldsymbol{\gamma}_0\right)\left(\hat{\boldsymbol{\gamma}}-\boldsymbol{\gamma}_0\right)\xrightarrow[]{d} N\left(\textbf{0},\textbf{I}\right)$.
Apply corollary 3 in \citep{KAUFMANN:1985}, we have Lemma 1.

\subsection*{ Proof of the Theorem 2}
We assume conditions in Lemma 1 and Assumption 1 are satisfied. 

Under $H_0$,  $\hat{{\Lambda}}_C$  asymptotically follows a multinormal distribution given by
$N_{G-1}\left(\textbf{0}_{G-1},\Sigma_{G-1}\right)$ and under the alternative, $\hat{{\Lambda}}_C$ follows $N_{G-1}\left(\hat{{\Lambda}}_{0C},\Sigma_{G-1}\right)$ as $ N \xrightarrow{} \infty$.

Applying Theorem 3.5 in \citep{serfling2009approximation} with $C=\Sigma_{G-1}^{-1}$, under $H_0$, $\Psi=\widehat{\boldsymbol\Lambda}_{C}\Sigma_{G-1}^{-1}\widehat{\boldsymbol\Lambda}_{C}^{\prime}$ has a chi-squared distribution with the degrees of freedom to be equal $trace\left(C\Sigma_{G-1}\right)=G-1$. Similarly, under $H_a$, $\Psi$ follows a noncentral chi-squared distribution  with a non-centrality parameter $\eta={\boldsymbol\Lambda}_{0C}\Sigma_{G-1}^{-1}{\boldsymbol\Lambda}_{0C}^{\prime}$.   
 \newpage
\subsection*{Post hoc Pairwise Comparisons of ROC curves}

\begin{figure}[!htp]
  \begin{center}
\includegraphics[height = .4\textheight]{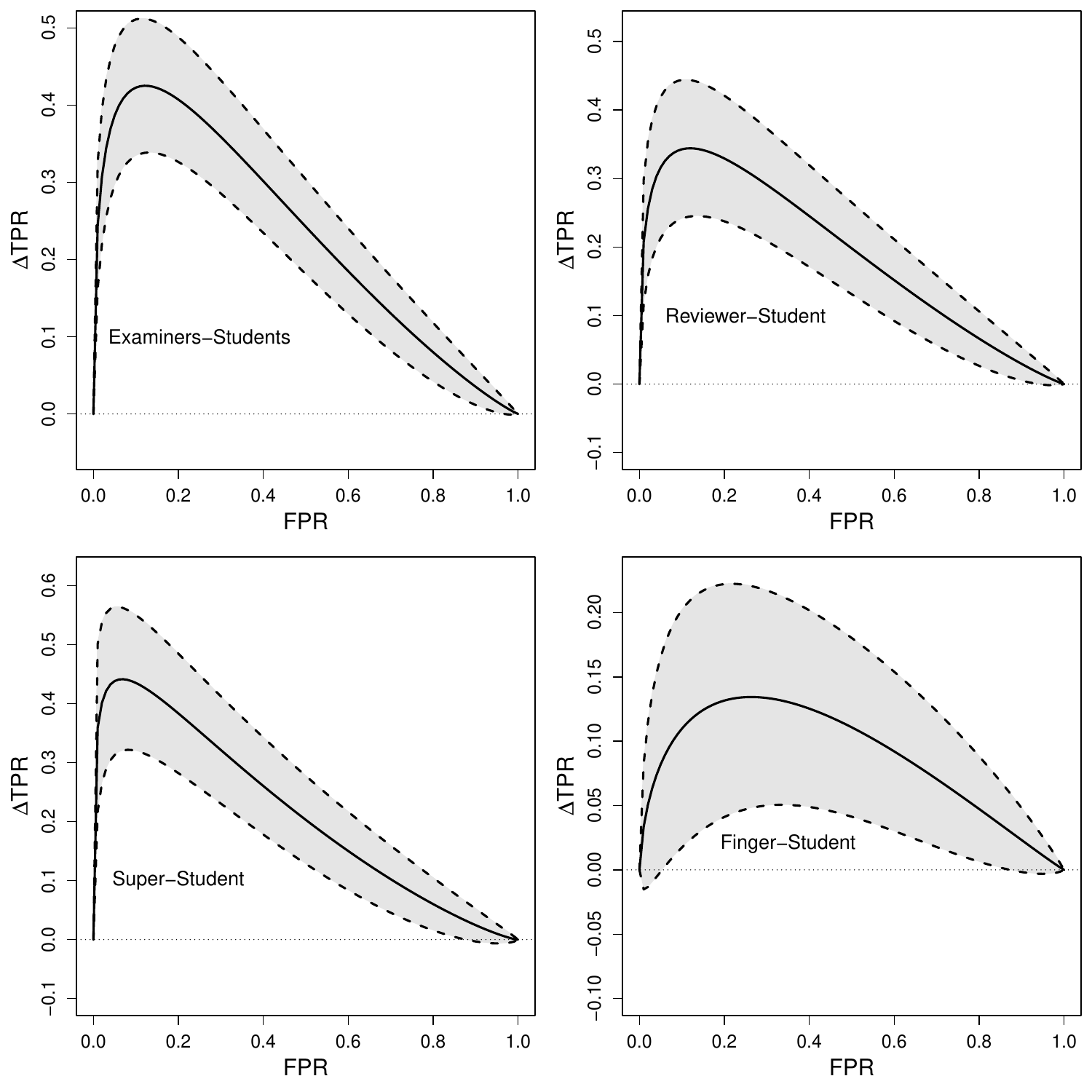}
\end{center}
\caption{Differences in ROC curves of five groups in facial recognition data. The solid line is estimated difference. Dashed lines are upper and lower bounds of the $95\%$ confidence band.  }
\label{fig:delROC2}
\end{figure}

\end{document}